\documentclass{mn2e}
\usepackage{graphics}
\usepackage{epsfig}

\def\ni{\noindent}                                       
\def\etal{et\thinspace al.\ }                               

\title[The star formation history of Seyfert 2 nuclei]
       {The star formation history of Seyfert 2 nuclei}

\author[Cid Fernandes, Gu, Melnick, Terlevich, Terlevich, Kunth,
        Rodrigues Lacerda, Joguet] {R. Cid
        Fernandes$^{1}$\thanks{E-mail: cid@astro.ufsc.br},
        Q. Gu$^{2}$\thanks{E-mail: qsgu@nju.edu.cn},
        J. Melnick$^{3}$\thanks{E-mail: jmelnick@eso.org},
        E. Terlevich$^{4}$\thanks{E-mail: eterlevi@inaoep.mx. Visiting
        fellow, Institute of Astronomy, Cambridge, U.K.},
        R. Terlevich$^{4}$\thanks{E-mail: rjt@ast.cam.ac.uk. Visiting
        fellow, Institute of Astronomy, Cambridge, U.K.},
	 \newauthor
 	 D. Kunth$^{5}$\thanks{E-mail: kunth@iap.fr},
	 R. Rodrigues Lacerda$^{1}$\thanks{E-mail: reiner@fsc.ufsc.br},
	 B. Joguet$^{5}$ \\
	 $^{1}$Departamento de F\'{\i}sica - CFM - Universidade Federal de Santa Catarina, PO Box 476, 
	 Florian\'opolis 88040-900, SC, Brazil\\
	 $^{2}$Department of Astronomy, Nanjing University, Nanjing
	 210093, P. R. China\\
	 $^{3}$European Southern Observatory, Alonso de Cordova 3107, Santiago, Chile\\ 
	 $^{4}$INAOE, Tonantzintla, Puebla, M{\'e}xico\\
	 $^{5}$Institut d'Astrophysique de Paris, 98bis Boulevard Arago, 75014
	 Paris, France}

\begin{document}

\maketitle

\begin{abstract} 
We present a study of the stellar populations in the central $\sim
200$ pc of a large and homogeneous sample comprising 79 nearby
galaxies, most of which are type 2 Seyferts. The star-formation
history of these nuclei is reconstructed by means of state-of-the art
population synthesis modeling of their spectra in the 3500--5200 \AA\
interval. A QSO-like featureless continuum (FC) is added to the models
to account for possible scattered light from a hidden AGN.

We find that: (1) The star-formation history of Seyfert~2 nuclei is
remarkably heterogeneous: young starbursts, intermediate age, and old
stellar populations all appear in significant and widely varying
proportions. (2) A significant fraction of the nuclei show a strong FC
component, but this FC is not always an indication of a hidden AGN: it
can also betray the presence of a young, dusty starburst. (3) We
detect weak broad H$\beta$ emission in several Seyfert 2s after
cleaning the observed spectrum by subtracting the synthesis
model. These are most likely the weak scattered lines from the hidden
Broad Line Region envisaged in the unified model, given that in most of
these cases independent spectropolarimetry data finds a hidden Seyfert
1. (4) The FC strengths obtained by the spectral decomposition are
substantially larger for the Seyfert 2s which present evidence of
broad lines, implying that the scattered non-stellar continuum is also
detected. (5) There is no correlation between the star-formation in
the nucleus and either the central or overall morphology of the parent
galaxies.
\end{abstract}

\begin{keywords} galaxies: active - galaxies: Seyfert - galaxies: 
stellar content - galaxies: statistics
\end{keywords}

\section{Introduction}

Our understanding of how galaxies form and evolve has changed in a
rather unexpected way in recent years. On the one hand, massive
black-holes now appear to populate the nuclei of virtually every
(massive enough) galaxy. On the other, the nuclei of active galaxies
(AGN), previously thought to be the only galaxies that harbored
massive black-holes, are now known to be the hosts of massive
star-forming regions. Hence, black-holes and starburst clusters coexist
and are ubiquitous in the nuclear regions of galaxies.

The observational evidence for this scenario is abundant.  The
presence of strong CaII triplet absorptions in a large sample of
Seyfert~2s and some Seyfert~1s provided the first direct evidence for
a population of red super-giant stars in their nuclear regions
(Terlevich, D\'{\i}az \& Terlevich 1990).  The absence of signs of a
Broad Line Region (BLR) in the direct optical spectra of Seyfert~2
nuclei which show broad lines in polarized light can only be
understood if there is an additional central source of blue/UV
continuum associated with the obscuring torus (Cid Fernandes \&
Terlevich 1995).  This conclusion is further supported by observations
that polarization is lower in the continuum than in the scattered
broad lines (Miller \& Goodrich, 1990; Tran 1995a,b)

Most of the UV to near-IR continuum light in a number of Seyfert~2s,
in particular all those so far observed with HST, is due to nuclear
starbursts that are resolved in HST images. In some of these galaxies,
the starburst luminosities are comparable to those of their ionizing
engines.  Thus, an observer situated along the axis of the torus will
detect comparable contributions to the optical continuum coming from
the starburst and from the broad line region components (Heckman \etal
1995, 1997).  Massive star forming regions are even found in the
nuclei of powerful radio galaxies through the detection of Balmer
absorption lines or strong CaII triplet lines (Melnick \etal 1997;
Aretxaga \etal 2001; Wills \etal 2002), showing that the intimate
link between nuclear activity and star-formation exists not only in
spiral galaxies, but also in massive ellipticals.

We have been engaged in a large spectroscopic survey of the nearest
Seyfert galaxies in the southern hemisphere aimed at characterizing
the properties of nuclear star-forming regions as a first step to
understand the link between nuclear starbursts and massive black
holes.  The first results of this project were reported in Joguet
\etal (2001, hereafter J01), where high S/N observations of about 80
Seyfert~2 galaxies were presented. This first inspection of the data
revealed that about 50\% of Seyfert 2s exihibit clear spectral
signatures of star-formation in the recent past ($\la 100$ Myr), most
notably high-order Balmer lines in absorption. The other half of the
sample is equally split between nuclei with strong emission lines but
weak absorption lines and nuclei with a rich absorption line spectrum
and reddish continuum, characteristic of old stellar populations.
These results are in line with those obtained with smaller samples
(Schmitt, Storchi-Bergmann \& Cid Fernandes 1999; Gonz\'alez-Delgado,
Heckman \& Leitherer 2001; Cid Fernandes \etal 2001a, hereafter CF01),
and highlight the high frequency of star-formation in Seyfert 2s.

Recently we have also started to detect high order Balmer lines in
Seyfert 1s which may indicate the presence of extremely compact and
luminous starbursts.  The fraction of Seyfert~1 nuclei with Balmer
absorption is understandingly lower than in Seyfert 2s given the
additional difficulty of disentangling the stellar absorption features
from the broad emission lines in Seyfert 1s (Torres-Papaqui \etal
2004, in preparation).

This paper is the first of a series devoted to a quantitative analysis
of the stellar populations in the nuclei of nearby Seyfert
galaxies. We present detailed population synthesis models for a sample
of 79 galaxies studied by J01. For each galaxy we derive the nuclear
star-formation history (SFH), and correlate the nuclear stellar
populations with other properties such as the emission line spectra
and the (central) morphology of the parent galaxy. The models are also
used to produce continuum-subtracted ``pure emission line'' spectra
which will be used in subsequent papers to study other properties of
the nuclei including emission-line profiles, the relation between
stellar and nebular kinematics, etc. The ultimate aim of this study is
to have at least a first glimpse at the coveted evolutionary
connection between monsters and starbursts.

This first paper concentrates on the presentation of our synthesis
method and the results of its application to the J01 sample.  Section
\ref{sec:data} reviews the data set used in this study.  A spectral
synthesis method which combines tools from empirical population
synthesis with the new Bruzual \& Charlot (2003, hereafter BC03)
models is described in Section \ref{sec:SpecSynthesis} and its
application to 65 Seyfert 2s and 14 other galaxies is presented in
Section \ref{sec:J01Synthesis}. The model subtracted spectra are used
to search for weak broad emission features in Section
\ref{sec:Starlight_Subtracted_Spectra}.  Section \ref{sec:Indices}
presents an empirical characterization of the stellar populations in
the sample based on a set of spectral indices. The statistics of
star-formation in Seyfert 2s and relations to host morphology are
discussed in Section \ref{sec:Discussion}.  Finally, Section
\ref{sec:Conclusions} summarizes our conclusions.

\section{The data set}

\label{sec:data}

Highly efficient blue spectroscopic observations of nearby Seyfert 2
galaxies were carried out in four observing runs between May 1998 and
August 2000, with the 1.5m ESO telescope at La Silla (Chile). The
spectra cover a wavelength range from 3470 to 5450 \AA , with a
spatial sampling of $0.82^{\prime\prime}$ per pixel. A detailed
description of the acquisition and reduction of the data, as well as of
the selection of the sample has been published in J01.

For this work, 1D spectra have been extracted afresh restricting
ourselves to the central pixel, which, at the redshifts of the galaxy
sample and adopting $H_0 = 75$ km$\,$s$^{-1}$Mpc$^{-1}$, represents the
central 59 to 439 pc for the Seyfert 2 galaxies (median = 174 pc), and
45 to 596 pc for the Seyfert 1, LINER, starburst and normal ones
(median = 121 pc). Table \ref{tab:Sample} lists some properties of the
galaxies in the sample, including morphology, far-infrared properties
and the linear scale.

J01 listed 15 galaxies of the sample with dubious (or wrong) Seyfert 2
classification.  After inspection of the spectra and a revision of
literature data, six extra objects were reclassified.  NGC 1097
(Storchi-Bergmann \etal 2003), NGC 4303 (Colina \etal 2003;
Jim\'enez-Bail\'on \etal 2003) and NGC 4602 were classified as
LINERs. NGC 2935 and the mergers NGC 1487 and NGC 3256 were
reclassified as Starbursts (Dessauges-Zavadsky \etal 2000; Lira \etal
2002). After these minor revisions, the final sample has 65 Seyfert
2s, 1 Seyfert 1, 4 LINERs, 4 Starburst/HII nuclei and 5 normal
galaxies.  Note that our list of Seyfert 2s includes some mixed cases,
such as NGC 6221 (Levenson \etal 2001), NGC 7496 (J01) and 7679 (Gu
\etal 2001; Della Ceca \etal 2001). Except for subtle details like
high-excitation components in the wings of the [OIII] and H$\beta$
profiles, these nuclei look like starbursts in the optical but have
X-ray properties of AGN (see also Maiolino \etal 2003; Gon\c{c}alves,
V\'eron-Cetty \& V\'eron 1999).

\begin{table*}
\tiny
\begin{centering}
\begin{tabular}{lrrrrrrr}
\multicolumn{8}{c}{Sample Properties}\\ \hline
  Name	        & 
  Type	        &
  Hubble	&
  I-type	& 
pc/$^{\prime\prime}$ &
$\alpha$(25,60) &
$\alpha$(60,100)&
$\log L_{FIR}$  \\ \hline
  ESO 103-G35    & Sy2     & S0       &  E    & 257 &        0.02  &   1.54  &  9.87 \\
  ESO 104-G11    & Sy2     & SBb      &       & 293 &        -2.00  &  -1.22  & 10.43 \\
  ESO 137-G34    & Sy2     & SAB0/a   &  S0   & 178 &        -1.29  &  -4.31  & 10.12 \\
  ESO 138-G01    & Sy2     & E        &  L    & 177 &      -0.47  &   0.20  &  9.66 \\
  ESO 269-G12    & Sy2     & S0       &       & 320 &              &         &       \\
  ESO 323-G32    & Sy2     & SAB0     &       & 310 &       -1.35  &  -1.11  &  9.81 \\
  ESO 362-G08    & Sy2     & S0       &  Sa   & 309 &        -1.45  &  -1.67  &  9.68 \\
  ESO 373-G29    & Sy2     & SBab     &  Sb   & 181 &              &         &       \\
  ESO 381-G08    & Sy2     & SBbc     &  SBb  & 212 &       -1.30  &  -0.85  & 10.33 \\
  ESO 383-G18    & Sy2     & S0/a$^a$ &       & 240 &       -0.60  &  -0.81  &  9.40 \\
  ESO 428-G14    & Sy2     & S0/a     &       & 105 &      -1.04  &  -0.62  &  9.49 \\
  ESO 434-G40    & Sy2     & S0/a     &       & 160 &              &         &       \\
  Fair 0334      & Sy2     & SB       &  SBb  & 359 &       -1.15  &  -1.46  &  9.84 \\
  Fair 0341      & Sy2     & S0/a     &  Sa?  & 312 &       -1.05  &  -0.69  &  9.75 \\
  IC 1657        & Sy2     & SBbc     &       & 232 &       -2.66  &  -1.94  & 10.10 \\
  IC 2560        & Sy2     & SBb      &       & 189 &       -1.41  &  -1.24  &  9.92 \\
  IC 5063        & Sy2     & S0       &       & 220 &       -0.36  &   0.48  & 10.14 \\
  IC 5135        & Sy2     & Sa       &       & 313 &       -2.34  &  -0.86  & 11.03 \\
  IRAS 11215-2806& Sy2     & S0$^a$   &  S0   & 262 &       -0.59  &  -0.76  &  9.40 \\
  MCG +01-27-020 & Sy2     & ?        &       & 227 &       -0.51  &  -2.07  &  9.24 \\
  MCG -03-34-064 & Sy2     & SB       &       & 321 &       -0.83  &   0.14  & 10.53 \\
  Mrk 0897       & Sy2     & Scd      &       & 510 &       -1.95  &  -1.00  & 10.69 \\
  Mrk 1210       & Sy2     & Sa$^a$   &  Sa   & 261 &        0.11  &   0.73  &  9.83 \\
  Mrk 1370       & Sy2     & Sa$^a$   &  Sa   & 476 &              &         &       \\
  NGC 0424       & Sy2     & SB0/a    &  Sb   & 226 &       -0.04  &  -0.02  &  9.72 \\
  NGC 0788       & Sy2     & S0/a     &  S0   & 264 &        0.00  &  -0.29  &  9.32 \\
  NGC 1068       & Sy2     & Sb       &       &  73 &      -0.83  &  -0.47  & 10.77 \\
  NGC 1125       & Sy2     & SB0/a    &  Sb/c & 212 &       -1.58  &  -0.34  &  9.95 \\
  NGC 1667       & Sy2     & SABc     &  Sc   & 294 &       -2.48  &  -1.77  & 10.62 \\
  NGC 1672       & Sy2     & SBb      &       &  87 &      -2.40  &  -1.47  & 10.28 \\
  NGC 2110       & Sy2     & SAB0     &  Sa   & 151 &       -1.82  &  -0.62  &  9.77 \\
  NGC 2979       & Sy2     & Sa       &       & 176 &       -2.05  &  -1.77  &  9.56 \\
  NGC 2992       & Sy2     & Sa       &       & 149 &       -2.18  &  -2.43  & 10.26 \\
  NGC 3035       & Sy2     & SBbc     &       & 281 &       -1.47  &  -1.78  &  9.62 \\
  NGC 3081       & Sy2     & SAB0/a   & SB0/a & 154 &               &         &       \\
  NGC 3281       & Sy2     & Sab      &       & 207 &        -1.10  &  -0.18  & 10.24 \\
  NGC 3362       & Sy2     & SABc     &  Sb   & 536 &               &         &       \\
  NGC 3393       & Sy2     & SBa      &  Sa   & 242 &        -1.25  &  -1.06  &  9.96 \\
  NGC 3660       & Sy2     & SBbc     &       & 238 &        -2.44  &  -1.74  &  9.93 \\
  NGC 4388       & Sy2     & Sb       &       & 163 &        -1.23  &  -1.12  & 10.28 \\
  NGC 4507       & Sy2     & SABb     & SBa/b & 229 &        -1.29  &  -0.44  & 10.14 \\
  NGC 4903       & Sy2     & SBc      &       & 319 &        -1.12  &  -2.23  &  9.85 \\
  NGC 4939       & Sy2     & Sbc      &  Sa   & 201 &        -1.91  &  -2.61  &  9.92 \\
  NGC 4941       & Sy2     & SABab    &       &  72 &        -1.09  &  -2.17  &  8.80 \\
  NGC 4968       & Sy2     & SAB0/a   &  Sa   & 191 &        -0.93  &  -0.41  &  9.72 \\
  NGC 5135       & Sy2     & SBab     &  Sc   & 266 &        -2.23  &  -1.03  & 10.91 \\
  NGC 5252       & Sy2     & S0       &  S0   & 447 &               &         &       \\
  NGC 5427       & Sy2     & Sc       &  Sc   & 169 &        -2.55  &  -2.90  & 10.24 \\
  NGC 5506       & Sy2     & Sa       &       & 120 &        -0.96  &  -0.11  &  9.85 \\
  NGC 5643       & Sy2     & SABc     &       &  77 &        -1.91  &  -1.32  &  9.93 \\
  NGC 5674       & Sy2     & SABc     &  SBc  & 483 &        -1.87  &  -1.86  & 10.44 \\
  NGC 5728       & Sy2     & SABa     &       & 180 &        -2.54  &  -1.15  & 10.27 \\
  NGC 5953       & Sy2     & Sa       &  Sc   & 127 &        -2.46  &  -1.26  & 10.06 \\
  NGC 6221       & Sy2     & SBc      &  Sd   &  96 &        -2.20  &  -1.65  & 10.42 \\
  NGC 6300       & Sy2     & SBb      &  Sd   &  72 &        -2.13  &  -1.77  &  9.78 \\
  NGC 6890       & Sy2     & Sb       &       & 156 &        -2.03  &  -1.46  &  9.85 \\
  NGC 7172       & Sy2     & Sa       &  ?    & 168 &        -2.30  &  -1.50  & 10.09 \\
  NGC 7212       & Sy2     & S?       &  Irr? & 516 &        -1.51  &  -1.04  & 10.72 \\
  NGC 7314       & Sy2     & SABbc    &  Sd   &  92 &        -2.13  &  -2.60  &  9.51 \\
  NGC 7496       & Sy2     & SBb      &       & 107 &        -1.90  &  -1.20  &  9.83 \\
  NGC 7582       & Sy2     & SBab     &  ?    & 102 &        -2.32  &  -0.77  & 10.52 \\
  NGC 7590       & Sy2     & Sbc      &  Sd   & 103 &        -2.42  &  -2.41  &  9.85 \\
  NGC 7679       & Sy2     & SB0      &       & 332 &        -2.13  &  -0.56  & 10.71 \\
  NGC 7682       & Sy2     & SBab     &  SB0  & 332 &               &         &       \\
  NGC 7743       & Sy2     & SB0      &  S0   & 110 &        -1.57  &  -2.28  &  8.95 \\ \hline
  Mrk 0883       & Sy1     & ?        &       & 727 &        -1.75  &  -0.22  & 10.50 \\
  NGC 1097       & LINER   & E        &       &  82 &        -2.39  &  -1.27  & 10.34 \\
  NGC 4303       & LINER   & SABbc    &       & 101 &        -3.23  &  -1.97  & 10.31 \\
  NGC 4602       & LINER   & SABbc    &       & 164 &        -2.47  &  -1.86  & 10.03 \\
  NGC 7410       & LINER   & SBa      &  E/S0 & 113 &        -2.49  &  -2.80  &  8.99 \\
  ESO 108-G17    & STB     & I0$^a$   &       & 142 &        -2.20  &  -0.90  &  9.26 \\
  NGC 1487       & STB     & P        &       &  55 &        -2.89  &  -1.32  &  8.85 \\
  NGC 2935       & STB     & SABb     &       & 147 &        -2.55  &  -2.18  &  9.85 \\
  NGC 3256       & STB     & SBm      &       & 177 &        -1.95  &  -0.52  & 11.23 \\
  NGC 2811       & Normal  & SBa      &       & 153 &        -1.51  &  -2.67  &  9.04 \\
  NGC 3223       & Normal  & Sb       &       & 187 &        -2.49  &  -2.67  & 10.14 \\
  NGC 3358       & Normal  & SAB0/a   &       & 193 &               &         &       \\
  NGC 3379       & Normal  & E        &       &  59 &               &         &       \\
  NGC 4365       & Normal  & E        &       &  80 &               &         &       \\ \hline
\end{tabular}
\end{centering}
\caption{Column (2): Spectral class.  Column (3): Hubble type, taken
from RC3 except for objects marked with $^{a}$, whose morphology was
taken from NED. Column (4): Inner Hubble type, taken from Malkan,
Gorjian \& Tam (1998). Column (5): Linear scale. Columns (6) and (7):
IRAS colours between 25, 60 and 100 $\mu$m. Column (8): Far infrared
luminosity, in L$_\odot$.}
\label{tab:Sample}
\end{table*}

\section{Spectral Synthesis}

\label{sec:SpecSynthesis}

\subsection{The method}

\label{sec:SynthesisMethod}

The modeling of stellar populations in galaxies has recently undergone
a major improvement with the publication of high spectral resolution
evolutionary synthesis models by BC03. We have incorporated this new
library into a modified version of the population synthesis code
described by Cid Fernandes \etal (2001b). The code searches for the
linear combination of $N_\star$ Simple Stellar Populations (SSP), ie,
populations of same age ($t$) and metallicity ($Z$), which best
matches a given observed spectrum $O_\lambda$. The equation for a
model spectrum $M_\lambda$ is:

\begin{equation}
M_\lambda(\vec{x},M_{\lambda_0},A_V,v_\star,\sigma_\star) = M_{\lambda_0}
   \left[
   \sum_{j=1}^{N_\star} x_j b_{j,\lambda} r_\lambda
   \right]
   \otimes G(v_\star,\sigma_\star)
\end{equation}

\ni where 

\begin{itemize}

\item[(i)] $b_{j,\lambda} \equiv L_\lambda^{SSP}(t_j,Z_j) /
L_{\lambda_0}^{SSP}(t_j,Z_j)$ is the spectrum of the $j^{\rm th}$ SSP
normalized at $\lambda_0$. The $L_\lambda^{SSP}(t_j,Z_j)$ spectra are
taken from BC03 models computed with the STELIB library (Le Borgne
\etal 2003), Padova tracks, and Chabrier (2003) mass function.

\item[(ii)] $r_\lambda \equiv 10^{-0.4 (A_\lambda - A_{\lambda_0})}$
is the reddening term, with $A_\lambda$ as the extinction at
wavelength $\lambda$. Extinction is modeled as due to an uniform dust
screen, parametrized by the V-band extinction $A_V$. The Cardelli,
Clayton \& Mathis (1989) extinction law with $R_V = 3.1$ is
adopted. 

\item[(iii)] $\vec{x}$ is the {\it population vector}, whose
components $x_j$ ($j = 1\ldots N_\star$) represent the fractional
contribution of each SSP in the base to the total synthetic flux at
$\lambda_0$.  These flux fractions can be converted to a mass
fractions vector $\vec{\mu}$ using the model light-to-mass ratios at
$\lambda_0$.

\item[(iv)] $M_{\lambda_0}$, the synthetic flux at the normalization
wavelength, plays the role of a scaling parameter.

\item[(v)] $G(v_\star,\sigma_\star)$ is the line-of-sight stellar
velocity distribution, modeled as a Gaussian centered at velocity
$v_\star$ and broadened by $\sigma_\star$.

\end{itemize}

The match between model and observed spectra is evaluated by 

\begin{equation}
\label{eq:chi2}
\chi^2(\vec{x},M_{\lambda_0},A_V,v_\star,\sigma_\star) = 
   \sum_{\lambda=1}^{N_\lambda} 
   \left[
   \left(O_\lambda - M_\lambda \right) w_\lambda
   \right]^2
\end{equation}

\ni where the weight spectrum $w_\lambda$ is defined as the inverse of
the noise in $O_\lambda$. Regions containing emission lines or
spurious features (bad pixels or sky residuals) are masked out by
assigning $w_\lambda = 0$. 

We normalize $O_\lambda$ by its median value ($O_N$) in a window
containing $\lambda_0$ instead of normalizing it by $O_{\lambda_0}$,
which is obviously more sensitive to noise than the median flux. The
synthetic spectrum is thus modeled in units of $O_N$, and we expect to
obtain $M_{\lambda_0} \sim 1$ in these units. 

The search for the best parameters is carried out with a simulated
annealing method, which consists of a series of $N_M$
likelihood-guided Metropolis tours through the parameter space (see
Cid Fernandes \etal 2001b for an illustrated discussion of the
Metropolis method applied to the population synthesis problem). In
each iteration, the weights $w_\lambda$ are increased by a factor
$f_w$, thus narrowing the peaks in the
$\chi^2(\vec{x},M_{\lambda_0},A_V,v_\star,\sigma_\star)$ hyper-surface. The
step-size ($\epsilon$) in each parameter is concomitantly decreased
and the number of steps ($N_s$) is increased to insure that each
parameter can in principle random-walk through all its allowed range
of values (eg, $x_j = 0 \rightarrow 1$ for the $\vec{x}$
components). To speed up computations, the kinematical parameters
$v_\star$ and $\sigma_\star$ are kept fixed within the Metropolis
loops, and re-fit at the end of each loop by a simpler minimization
algorithm.

The output population vector usually contains many $x_j = 0$
components. The code offers the possibility of re-fitting the data
excluding these components, which in principle allows a finer search
for a best fit. In practice, however, this turns out to have little
effect upon the results reported here. We have also implemented a
conjugate-gradients routine which runs after each Metropolis loop, in
order to refine the search for a minimum $\chi^2$, but this feature
too turns out to be largely irrelevant.  The code ``clips'' deviant
points, assigning zero weight to pixels which are more than 3 sigma
away from the rms $O_\lambda - M_\lambda$ residual flux. This feature,
which obviously requires an initial estimate of $M_\lambda$, is useful
to mask weak emission lines or defects in $O_\lambda$ not eliminated
by the original $w_\lambda = 0$ mask.

This same general method, with variations on numerical schemes,
spectral base, extinction laws and other details has been used in
several recent investigations (eg, Kauffmann \etal 2003; Tremonti 2003;
Mayya \etal 2004; BC03).

\subsection{The spectral base and tests}

\label{sec:SynthesisTests}
\label{sec:GroupingSchemes}

A key ingredient in the synthesis of stellar populations is the
spectral base $b_{j,\lambda}$. Ideally, the elements of the base
should span the range of spectral properties observed in the sample
galaxies and provide enough resolution in age and metallicity to
address the desired scientific questions.

We used as a starting base the 30 SSPs used by Tremonti (2003) in her
study of SDSS galaxies. These SSPs cover 10 ages, $t = 5 \times 10^6$,
$2.5 \times 10^7$, $10^8$, $2.9 \times 10^8$, $6.4 \times 10^8$, $9
\times 10^8$, $1.4 \times 10^9$, $2.5 \times 10^9$, $5 \times 10^9$
and $1.1 \times 10^{10}$ yr, and 3 metallicities, $Z = 0.2$, 1 and
$2.5 Z_\odot$. Tremonti also included 9 other spectra in her base,
corresponding to continuous star formation models of different ages,
e-folding times and metallicities. Not surprisingly, we found that
these spectra are very well modeled by combinations of the SSPs in the
base. Their inclusion in the base therefore introduces unnecessary
redundancies, so we opted not to include these components. Smaller
bases, with these same ages but only one or two metallicities, were
also explored.

Extensive simulations were performed to evaluate the code's ability to
recover input parameters, to calibrate its technical parameters
(number of steps, step-size, etc.) and to investigate the effects of
noise in the data. To emulate the modeling of real galaxy spectra
discussed in \S\ref{sec:J01Synthesis}, the simulations were restricted
to the 3500--5200 \AA\ interval, with masks around the wavelengths of
strong emission lines.

\begin{figure*}
\resizebox{\textwidth}{!}{\includegraphics{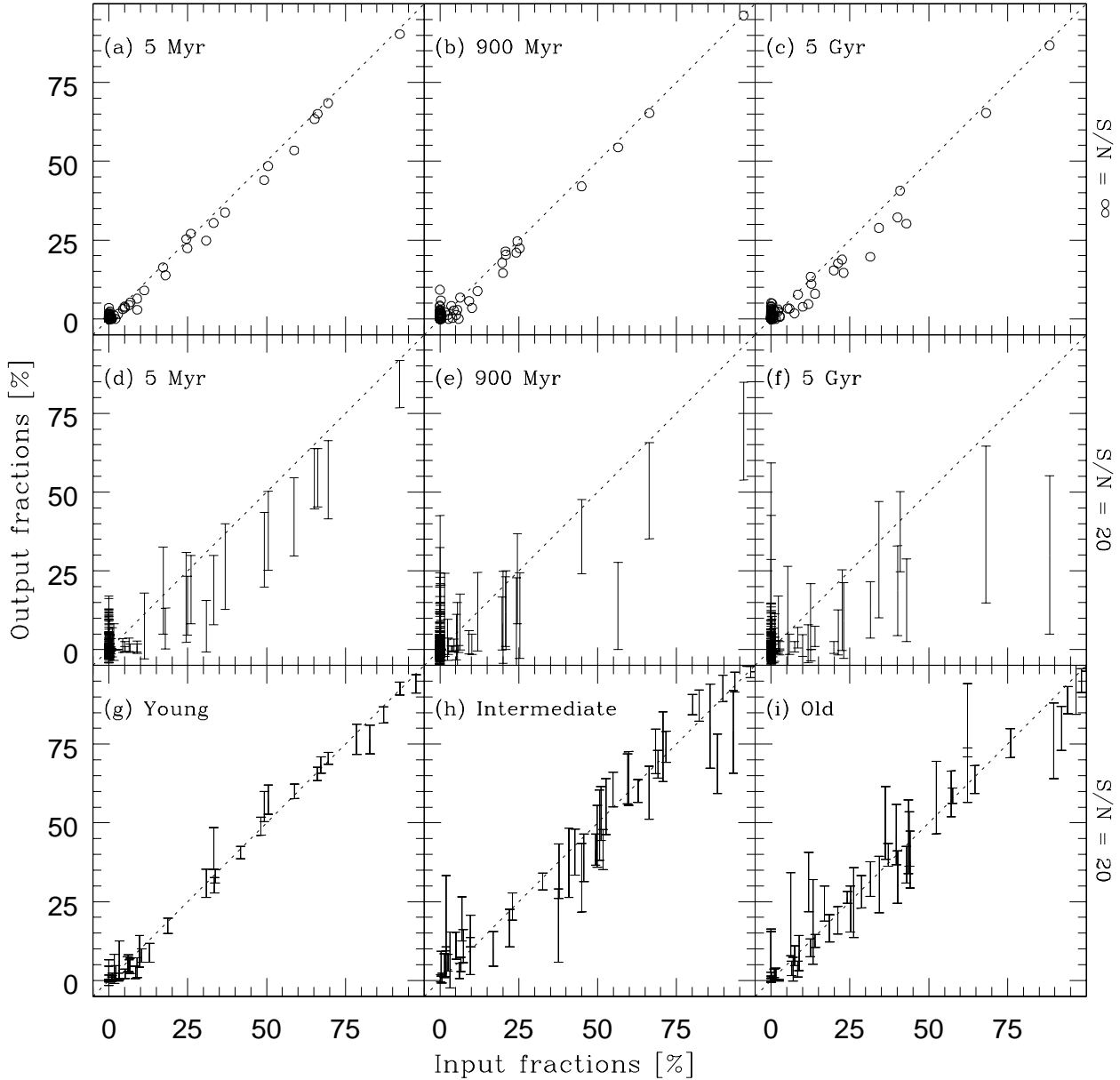}}
\caption{Tests of the spectral synthesis method. Panels a--c compare
input against output $x_j$ fractions for fake galaxies generated out
of the base and with no noise. Each panel contains individual $x_j$
components corresponding to 3 different metallicities but same age.
Panels d--f are equivalent to panels a--c, but for simulations with
noise of amplitude $S/N = 20$ added to each spectrum. The output
fractions are now represented by the mean $x_j$, and the error bars
correspond to the dispersion among 50 different realizations of the
noise for each galaxy.  The bottom panels (g--i) present the
population vector condensed into just three age groups, $x_Y$, $x_I$,
and $x_O$, binned-over all metallicities.  }
\label{fig:input_x_ouput}
\end{figure*}

Tests with fake galaxies generated out of the base show that in the
absence of noise the code is able to recover the input parameters to a
high degree of accuracy. This is illustrated in figures
\ref{fig:input_x_ouput}a--c, where we plot input against output $x_j$
fractions for 3 selected ages and all 3 metallicities. The rms
difference between input and output $x_j$ in these simulations is
typically $\sim 0.02$.

This good agreement goes away when noise is added to the spectra. As
found in other population synthesis methods (eg., Mayya \etal 2004),
noise has the effect of inducing substantial rearrangements among
$\vec{x}$ components corresponding to spectrally similar populations.
The reason for this is that noise washes away the differences between
$b_{j,\lambda}$ spectra which are similar due to intrinsic
degeneracies of stellar populations (like the age-metallicity
degeneracy; Worthey 1994). The net effect is that, for practical
purposes, the base becomes linearly dependent, such that a given
component is well reproduced by linear combinations of others, as
extensively discussed by Cid Fernandes \etal (2001b).  For realistic
$S/N$ ratios these rearrangements have disastrous effects upon the
individual $x_j$ fractions, as illustrated in figures
\ref{fig:input_x_ouput}d--f. The test galaxies in these plots are the
same ones used in the noise-free simulations (panels a--c), but with
Gaussian noise of amplitude $S/N = 20$ added to their spectra. Each
galaxy was modeled 50 times for different realizations of the noise,
and we plot the mean output $x_j$ with an error bar to indicate the
rms dispersion among these set of perturbed spectra. Clearly, the
individual $x_j$ fractions are not well recovered at all.  Hence,
while in principle it is desirable to use a large, fine graded base,
the resulting SFH expressed in $\vec{x}$ cannot be trusted to the same
level of detail.

A strategy to circumvent this problem is to {\it group} the $x_j$
fractions corresponding to spectrally similar components, which yields
much more reliable results at the expense of a coarser description of
the stellar population mixture.  This reduction can be done in a
number of ways (Pelat 1997; Moultaka \& Pelat 2000; Chan, Mitchell, \&
Cram 2003). Here we use our simulations to define combinations of
$\vec{x}$ components which are relatively immune to degeneracy
effects. We define a reduced population vector with three age ranges:
$t \le 25$ Myr; 100 Myr $\le t \le 1.4$ Gyr and $t \ge 2.5$ Gyr. The
resulting population vector is denoted by ($x_Y,x_I,x_O$), where $Y$,
$I$ and $O$ correspond to Young, Intermediate and Old
respectively. Note that the metallicity information is binned-over in
this description.  Figures \ref{fig:input_x_ouput}g--i show the input
against output ($x_Y,x_I,x_O$) components for the same $S/N= 20$
simulations described above. The deviations observed in the individual
$x_j$ fractions are negligible in this condensed description. 
Finally, we note that extinction does not suffer from the degeneracies
which plague the population vector. For $S/N = 20$ the output $A_V$
matches the input value to within 0.04 mag rms.

\section{Application to Seyfert 2s}

\label{sec:J01Synthesis}

The synthesis method described above was applied to the spectra
presented in \S\ref{sec:data} in order to estimate the star formation
history in the central regions of Seyfert 2s.

\subsection{Setup}

For the application described in this paper we have used the following
set-up: $N_M = 10$ Metropolis loops, with weights increasing
geometrically from $f_w = 0.1$ to 10 times their nominal values while
the step sizes decrease from $\epsilon = 0.1$ to 0.005 for both
the $\vec{x}$ components, $M_{\lambda_0}$ and $A_V$. The number of
steps in each Metropolis run is computed as a function of the number
of parameters being changed and by the step-size: $N_s = (N_\star + 1)
\epsilon^{-2}$.  Since we are modeling galactic nuclei, it is
reasonable to neglect the contribution of metal poor populations. We
thus opted to use a base of $N_\star = 20$ components with the same 10
ages described above but reduced to $Z = Z_\odot$ and $2.5 Z_\odot$.
This setup gives a total of $\sim 3 \times 10^6$ steps in the search
for a best model.

\subsection{Pre-processing}

Prior to the synthesis all spectra were corrected for Galactic
extinction using the law of Cardelli \etal (1989) and the $A_B$ values
from Schlegel, Finkbeiner \& Davis (1998) as listed in
NED\footnote{The NASA/IPAC Extragalactic Database (NED) is operated by
the Jet Propulsion Laboratory, California Institute of Technology,
under contract with the National Aeronautics and Space
Administration.}.  The spectra were also re-binned to 1 \AA\ sampling
and shifted to the rest-frame. The fitted values of $v_\star$ should
therefore be close to zero.

We chose to normalize the base at $\lambda_0 = 4020$ \AA, while the
observed spectra are normalized to the median flux between 4010 and
4060 \AA.  This facilitates comparison with previous studies which
used a similar normalization (eg., Gonz\'alez Delgado \etal 2004).  A
flat error spectrum is assumed.  We estimate the error in $O_\lambda$
from the rms flux in the relatively clean window between 4760 and 4810
\AA.  The signal-to-noise ratio varies between $S/N \sim 12$ and 67,
with a median of 27 in this interval, and about half these values in
the 4010--4060 \AA\ range.

Masks of 20--30 \AA\ around obvious emission lines were constructed
for each object individually. In sources like NGC 3256 and ESO
323-G32, these $w_\lambda = 0$ masks remove relatively small portions
of the original spectra. In several cases, however, the emission line
spectrum is so rich that the unmasked spectrum contains relatively few
stellar features apart from the continuum shape (eg, Mrk 1210, NGC
4507). It some cases it was also necessary to mask the $\lambda <
3650$ \AA\ region because of strong nebular emission in the Balmer
continuum (eg, NGC 7212). The strongest stellar features which are
less affected by emission are Ca II K and the G-band. We have
increased the weight of these two features by multiplying $w_\lambda$
by 2 in windows of $\sim 40$ \AA\ around their central wavelengths,
although this is of little consequence for the results reported in
this paper.

\subsection{Featureless Continuum component}

\label{sec:FC_component}

A non-stellar component, represented by a $F_\nu \propto \nu^{-1.5}$
power-law, was added to the stellar base to represent the contribution
of an AGN featureless continuum, a traditional ingredient in the
spectral modeling of Seyfert galaxies since Koski (1978).  The
fractional contribution of this component to the flux at $\lambda_0$
is denoted by $x_{FC}$.

In the framework of the unified model (eg, Antonucci 1993), in Seyfert
2s this FC, if present, must be scattered radiation from the hidden
Seyfert 1 nucleus.  Nevertheless, one must bear in mind that young
starbursts can easily appear disguised as an AGN-looking continuum in
optical spectra, a common problem faced in spectral synthesis of
Seyfert 2s (Cid Fernandes \& Terlevich 1995; Storchi-Bergmann \etal
2000). The interpretation of $x_{FC}$ thus demands care. This issue is
further discussed in \S\ref{sec:FC_X_YoungStars}.

\subsection{Spectral Synthesis Results}

\label{sec:J01Synthesis_Results}

\begin{table*}
\tiny
\begin{centering}
\begin{tabular}{lrrrrrrrrrrrl}
\multicolumn{13}{c}{Synthesis Results}\\ \hline
Galaxy           &
$x_{FC}$         &
$x_Y$            &   
$x_I$            & 
$x_O$            & 
$\mu_Y$          & 
$\mu_I$          & 
$\mu_O$          & 
$A_V$            &
$\sigma_\star$   &
$\chi^2_\lambda$ &
$\Delta_\lambda$ &
Notes            \\ \hline
ESO 103-G35       &    3  &    4  &    0  &   93    &   0.00 &   0.00 & 100.00  &   0.36  &   114   &     0.9 &    4.3 \\
ESO 104-G11       &    3  &   35  &   32  &   30    &   0.07 &   8.25 &  91.68  &   1.30  &   130   &     1.5 &   12.7 \\
ESO 137-G34       &   15  &    0  &    0  &   85    &   0.00 &   0.00 & 100.00  &   0.18  &   133   &     1.6 &    4.1 \\
ESO 138-G01       &    4  &   49  &   35  &   12    &   0.32 &  48.74 &  50.95  &   0.00  &    80   &     2.6 &    2.8 & WR?\\
ESO 269-G12       &    0  &    1  &   11  &   88    &   0.00 &   0.52 &  99.48  &   0.00  &   161   &     1.4 &    6.5 \\
ESO 323-G32       &    0  &    0  &   29  &   71    &   0.00 &   1.48 &  98.52  &   0.37  &   131   &     0.5 &    4.1 \\
ESO 362-G08       &    0  &    0  &   96  &    4    &   0.00 &  30.38 &  69.62  &   0.36  &   154   &     1.4 &    3.0 \\
ESO 373-G29       &   27  &   22  &   29  &   21    &   0.06 &   6.97 &  92.97  &   0.03  &    92   &     1.6 &    3.3 \\
ESO 381-G08       &   24  &    3  &   67  &    6    &   0.01 &  30.08 &  69.91  &   0.00  &   100   &     0.7 &    2.1 & BLR?, HBLR \\
ESO 383-G18       &   31  &   24  &    0  &   45    &   0.06 &   0.00 &  99.94  &   0.00  &    92   &     1.5 &    5.0 &  BLR \\
ESO 428G14        &    0  &   20  &   47  &   33    &   0.03 &  13.23 &  86.75  &   0.53  &   120   &     0.6 &    3.7 &  n-HBLR \\
ESO 434G40        &    0  &   10  &   11  &   79    &   0.00 &   0.62 &  99.38  &   0.39  &   145   &     0.3 &    3.9 &  HBLR \\
Fai 0334          &    0  &   22  &   47  &   31    &   0.03 &  14.72 &  85.25  &   0.31  &   104   &     1.1 &    7.8 \\
Fai 0341          &   13  &    0  &    1  &   86    &   0.00 &   0.23 &  99.77  &   0.26  &   122   &     1.7 &    8.4 \\
IC 1657           &    0  &    6  &   33  &   61    &   0.01 &   4.98 &  95.01  &   0.51  &   143   &     1.1 &    8.4 \\
IC 2560           &    0  &   33  &    0  &   67    &   0.02 &   0.00 &  99.98  &   0.47  &   144   &     1.4 &    4.1 \\
IC 5063           &    6  &   16  &    0  &   78    &   0.01 &   0.00 &  99.99  &   0.57  &   182   &     1.2 &    5.5 &  HBLR \\
IC 5135           &   35  &   34  &   31  &    0    &   0.74 &  99.26 &   0.00  &   0.00  &   143   &     2.6 &    2.0 \\
IRAS 11215-2806   &    2  &    8  &   33  &   56    &   0.02 &   9.54 &  90.45  &   0.55  &    98   &     1.7 &    5.9 \\
MCG +01-27-020    &    0  &   20  &   64  &   15    &   0.06 &  16.55 &  83.39  &   0.70  &    94   &     1.1 &    5.3 \\
MCG -03-34-064    &    5  &   26  &    0  &   69    &   0.02 &   0.00 &  99.98  &   0.38  &   155   &     1.5 &    3.8 &  BLR, HBLR, WR \\
MRK 0897          &   18  &   19  &   60  &    3    &   0.20 &  60.56 &  39.24  &   0.00  &   133   &     1.8 &    2.1 \\
MRK 1210          &    3  &   53  &    5  &   39    &   0.07 &   0.89 &  99.05  &   1.03  &   114   &     4.1 &    5.9 &  HBLR, WR \\
MRK 1370          &    0  &    0  &   73  &   27    &   0.00 &  13.17 &  86.83  &   0.47  &    86   &     1.6 &    4.7 \\
NGC 0424          &   36  &   13  &    0  &   52    &   0.02 &   0.00 &  99.98  &   0.62  &   143   &     1.8 &    4.2 &  BLR, HBLR, WR \\
NGC 0788          &    0  &   13  &    0  &   87    &   0.01 &   0.00 &  99.99  &   0.00  &   163   &     1.3 &    6.0 &  HBLR \\
NGC 1068          &   23  &   24  &    0  &   54    &   0.04 &   0.00 &  99.96  &   0.00  &   144   &     1.1 &    3.1 &  BLR, HBLR \\
NGC 1125          &   10  &    0  &   64  &   26    &   0.00 &  48.92 &  51.08  &   0.04  &   105   &     1.6 &    5.8 \\
NGC 1667          &    0  &    8  &   26  &   66    &   0.01 &   2.54 &  97.45  &   0.58  &   149   &     1.6 &    8.9 & n-HBLR \\
NGC 1672          &    9  &   12  &   68  &   11    &   0.03 &  24.55 &  75.42  &   0.24  &    97   &     0.7 &    2.7 \\
NGC 2110          &    0  &   26  &    7  &   67    &   0.01 &   0.05 &  99.93  &   0.38  &   242   &     2.8 &    9.2 \\
NGC 2979          &    0  &    1  &   78  &   21    &   0.00 &  15.70 &  84.30  &   0.70  &   112   &     1.4 &    5.5 \\
NGC 2992          &    0  &    6  &   35  &   59    &   0.00 &   2.60 &  97.40  &   1.24  &   172   &     0.9 &    6.3 &  HBLR \\
NGC 3035          &    9  &    5  &   18  &   69    &   0.00 &   0.18 &  99.81  &   0.08  &   161   &     0.8 &    4.6 &  BLR \\
NGC 3081          &   21  &    5  &    0  &   74    &   0.00 &   0.00 & 100.00  &   0.06  &   134   &     0.5 &    3.0 &  HBLR \\
NGC 3281          &    2  &   16  &    0  &   82    &   0.01 &   0.00 &  99.99  &   0.67  &   160   &     0.8 &    5.6 &  n-HBLR \\
NGC 3362          &   12  &    0  &   30  &   58    &   0.00 &  10.39 &  89.61  &   0.21  &   104   &     1.2 &    5.7 &  n-HBLR \\
NGC 3393          &   14  &    4  &    0  &   83    &   0.00 &   0.00 & 100.00  &   0.06  &   157   &     0.4 &    2.7 \\
NGC 3660          &   45  &    0  &   28  &   26    &   0.00 &   3.98 &  96.02  &   0.06  &    95   &     1.0 &    2.9 &  BLR, n-HBLR \\
NGC 4388          &   16  &   13  &    0  &   70    &   0.02 &   0.00 &  99.98  &   0.73  &   111   &     1.5 &    4.3 &  HBLR \\
NGC 4507          &   24  &   19  &    0  &   57    &   0.02 &   0.00 &  99.98  &   0.00  &   144   &     0.6 &    2.2 &  BLR?, HBLR, WR? \\
NGC 4903          &    0  &    6  &   11  &   83    &   0.00 &   0.50 &  99.50  &   0.55  &   200   &     1.0 &    6.8 \\
NGC 4939          &    0  &   15  &    0  &   85    &   0.01 &   0.00 &  99.99  &   0.18  &   155   &     0.7 &    3.7 \\
NGC 4941          &    2  &    1  &   11  &   86    &   0.00 &   0.35 &  99.65  &   0.16  &    98   &     0.1 &    2.9 &  n-HBLR \\
NGC 4968          &    1  &   20  &   58  &   22    &   0.04 &  15.17 &  84.79  &   1.04  &   121   &     1.5 &    7.9 \\
NGC 5135          &   12  &   57  &   24  &    6    &   0.17 &   6.27 &  93.56  &   0.38  &   143   &     2.8 &    2.8 &  n-HBLR \\
NGC 5252          &    0  &   14  &    0  &   86    &   0.01 &   0.00 &  99.99  &   0.57  &   209   &     0.9 &    3.8 &  HBLR \\
NGC 5427          &    0  &   16  &   12  &   72    &   0.01 &   1.63 &  98.35  &   0.45  &   100   &     1.5 &    6.9 \\
NGC 5506          &   63  &    0  &    5  &   33    &   0.00 &   4.75 &  95.25  &   0.76  &    98   &     1.9 &    3.9 &  BLR, HBLR \\
NGC 5643          &    6  &    0  &   88  &    6    &   0.00 &  43.15 &  56.85  &   0.30  &    93   &     0.7 &    2.1 &  n-HBLR\\
NGC 5674          &    2  &    0  &   36  &   63    &   0.00 &   0.71 &  99.29  &   0.64  &   129   &     1.1 &   10.0 \\
NGC 5728          &    0  &   20  &   51  &   28    &   0.02 &   6.99 &  92.99  &   0.25  &   155   &     0.5 &    2.0 &  n-HBLR \\
NGC 5953          &   18  &    1  &   74  &    7    &   0.00 &  57.94 &  42.06  &   0.67  &    93   &     0.9 &    2.7 \\
NGC 6221          &   32  &   28  &   32  &    8    &   0.14 &  14.43 &  85.43  &   0.49  &   111   &     1.7 &    3.3 \\
NGC 6300          &    0  &    0  &   75  &   25    &   0.00 &  16.03 &  83.97  &   0.92  &   100   &     0.5 &    4.2 &  n-HBLR \\
NGC 6890          &   13  &    6  &    9  &   72    &   0.01 &   2.27 &  97.72  &   0.45  &   109   &     0.7 &    3.9 &  n-HBLR \\
NGC 7172          &    0  &   11  &   13  &   76    &   0.01 &   0.95 &  99.05  &   0.80  &   190   &     1.7 &   11.5 &  n-HBLR \\
NGC 7212          &   35  &    0  &    0  &   65    &   0.00 &   0.00 & 100.00  &   0.36  &   168   &     1.5 &    3.2 &  BLR, HBLR \\
NGC 7314          &   27  &    5  &    0  &   68    &   0.00 &   0.00 & 100.00  &   0.83  &    60   &     2.0 &   16.6 &  HBLR \\
NGC 7496          &   27  &   23  &   44  &    6    &   0.07 &   9.11 &  90.82  &   0.00  &   101   &     1.9 &    2.3 &  n-HBLR \\
NGC 7582          &   31  &    0  &   69  &    0    &   0.00 & 100.00 &   0.00  &   1.23  &   132   &     1.9 &    2.1 &  n-HBLR \\
NGC 7590          &    1  &    0  &   49  &   50    &   0.00 &   7.63 &  92.38  &   0.70  &    99   &     0.4 &    3.0 &  n-HBLR \\
NGC 7679          &    9  &    0  &   91  &    0    &   0.00 &  81.81 &  18.19  &   0.42  &    96   &     2.7 &    2.6 \\
NGC 7682          &    6  &    8  &    0  &   86    &   0.00 &   0.00 & 100.00  &   0.00  &   152   &     0.9 &    4.2 &  HBLR \\
NGC 7743          &    0  &    1  &   77  &   21    &   0.00 &  23.10 &  76.89  &   0.48  &    95   &     0.6 &    3.4 \\
MRK 0883          &   48  &   16  &   31  &    5    &   0.14 &  74.38 &  25.47  &   0.69  &   202   &     0.7 &    3.6 \\ \hline
NGC 1097          &   19  &   31  &   37  &   12    &   0.05 &   4.10 &  95.85  &   0.24  &   150   &     2.2 &    3.9 \\
NGC 4303          &   11  &    4  &   59  &   26    &   0.00 &  12.69 &  87.30  &   0.10  &    82   &     0.3 &    2.8 \\
NGC 4602          &    0  &   25  &   49  &   25    &   0.04 &   8.10 &  91.87  &   0.36  &    94   &     1.1 &    5.3 \\
NGC 7410          &   12  &   57  &   24  &    6    &   0.17 &   6.33 &  93.50  &   0.54  &   140   &     2.7 &    2.8 \\
ESO 108-G17       &   13  &   72  &   15  &    0    &   4.70 &  95.30 &   0.00  &   0.00  &   108   &    10.6 &    2.1 & WR \\
NGC 1487          &    0  &   96  &    0  &    4    &   0.75 &   0.00 &  99.25  &   0.00  &   145   &     6.0 &    5.4 \\
NGC 2935          &    0  &   12  &   38  &   50    &   0.01 &   1.93 &  98.07  &   0.37  &   143   &     0.5 &    3.2 \\
NGC 3256          &   24  &   45  &   30  &    0    &   1.00 &  47.94 &  51.07  &   0.45  &   128   &     4.7 &    2.5 \\
NGC 2811          &    0  &    0  &   13  &   87    &   0.00 &   0.08 &  99.92  &   0.28  &   246   &     0.6 &    4.2 \\
NGC 3223          &    0  &    0  &   12  &   88    &   0.00 &   0.15 &  99.85  &   0.18  &   170   &     0.7 &    5.7 \\
NGC 3358          &    0  &    0  &   18  &   82    &   0.00 &   0.56 &  99.44  &   0.21  &   194   &     0.7 &    5.0 \\
NGC 3379          &    0  &    0  &   14  &   86    &   0.00 &   0.27 &  99.73  &   0.00  &   217   &     0.7 &    3.8 \\
NGC 4365          &    0  &    0  &   17  &   83    &   0.00 &   0.13 &  99.87  &   0.29  &   257   &     0.7 &    4.6 \\ \hline
\end{tabular}
\end{centering}
\caption{Columns 2--5: Population vector in the $(x_{FC},x_Y,x_I,x_O)$
description, in percentage of the flux at 4020 \AA. Columns 6--8:
percentage mass fractions associated to $Y$, $I$ and $O$
components. Column 9: Extinction (in V-band magnitudes). Column 10:
Velocity dispersion, in km$\,$s$^{-1}$.  Column 11: $\chi^2$ per
synthesized $\lambda$. Column 12: Average absolute difference between
model and observed spectra, in percentage. Column 13: BLR = weak broad
H$\beta$; HBLR = hidden broad line region detected through
spectropolarimetry; n-HBLR = hidden broad line region not detected in
spectropolarimetry; WR = Wolf-Rayet features. Question marks indicate
dubious detections.}
\label{tab:synthesis}
\end{table*}

Figures \ref{fig:Fit_j001}--\ref{fig:Fit_j003} illustrate some of the
spectral fits. The examples were chosen to represent the variety of
spectra and star-formation histories found in the sample.  The figures
show the observed and synthetic spectra, as well as the $O_\lambda -
M_\lambda$ ``pure emission'' residual spectrum. Emission lines,
particularly weak ones, appear much more clearly in the residual
spectrum. The starlight subtraction also enhances the Balmer lines,
particularly in nuclei with a significant intermediate age stellar
population (eg, ESO362-G08). A detailed analysis of the emission lines
is postponed to a future communication.

The SFH plots (right panels in figures
\ref{fig:Fit_j001}--\ref{fig:Fit_j003}) show flux ($x_j$) and mass
($\mu_j$) fractions for all 10 ages spanned by the base, plus the FC
component.  Note that by concentrating on the age distribution we are
ignoring the metallicity information in $\vec{x}$. Even binning over
$Z$, a description of the SFH in terms of 10 age bins is too detailed
given the effects of noise, intrinsic degeneracies of the synthesis
process, and limited spectral coverage of the data which is even
further limited by the masks around emission lines. Rearrangements of
the $x_j$ strengths among adjacent age bins in these figures would
produce fits of comparable quality, as found in independent population
synthesis studies (eg, Mayya \etal 2004).

As discussed in \S\ref{sec:GroupingSchemes}, a coarser but more robust
description of the SFH requires further binning of the age
distributions in figures \ref{fig:Fit_j001}--\ref{fig:Fit_j008}. In
Table~\ref{tab:synthesis} we condense the population vector $\vec{x}$
onto just four components: an FC ($x_{FC}$) and three stellar
components representing young ($t \le 2.5 \times 10^7$ yr),
intermediate age ($10^8 \le t \le 1.4 \times 10^9$ yr), 
and old ($t \ge 2.5 \times 10^9$ yr) populations, denoted by
$x_Y$, $x_I$ and $x_O$ respectively. We note that the 25 Myr component
is zero or close to it in practically all galaxies, so that $x_Y$ is
virtually identical to the 5 Myr component. Mass fractions are
described into just three components: $\mu_Y$, $\mu_I$ and $\mu_O$,
since we do not associate a mass to the FC.

\begin{figure}
\includegraphics[width=9cm]{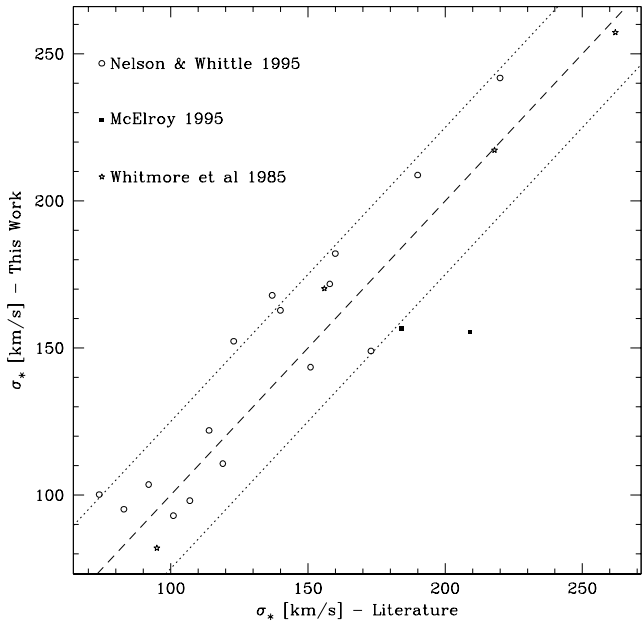}
\caption{Comparison of the stellar velocity dispersions estimated by
our synthesis method and values compiled from the literature. The
identity line is traced by the dashed line, while dotted lines
indicate the $\pm 1$ sigma dispersion.}
\label{fig:vd_US_X_Literature}
\end{figure}

The values of $A_V$ and $\sigma_\star$ produced by the synthesis are
also listed in Table~\ref{tab:synthesis}. The velocity dispersion was
corrected by the instrumental resolutions of both the J01 spectra
($\sigma_{inst} = 62$ km$\,$s$^{-1}$) and the STELIB library ($\sim
86$ km$\,$s$^{-1}$). Figure \ref{fig:vd_US_X_Literature} compares our
estimates of $\sigma_\star$ with values compiled from the literature
(mostly Nelson \& Whittle 1995). The agreement is good, with a mean
and rms difference of $3 \pm 21$ km$\,$s$^{-1}$. In the median,
our values are 12 km$\,$s$^{-1}$ larger than those in the literature,
possibly due to filling of stellar absorptions by weak emission lines
not masked out in the fits. On the whole, however, this is a minor
effect given that the uncertainty in $\sigma_\star$ is typically 20
km$\,$s$^{-1}$ both for our and the literature values.

The quality of the fits can be measured by $\chi^2_\lambda$,
which is the $\chi^2$ of equation (\ref{eq:chi2}) divided by the
effective number of wavelengths used in the synthesis (ie, discounting
the masked points). In most cases we obtain $\chi^2_\lambda \sim 1$,
indicative of a good fit. However, the value of $\chi^2_\lambda$
depends on the assumed noise amplitude and spectrum, as well as on
extra weights given to special windows, so standard $\chi^2$
statistical goodness-of-fit diagnostics do not apply. An alternative,
albeit rather informal, measure of the quality of the fits is given by
the mean absolute percentage deviation between $O_\lambda$ and
$M_\lambda$ for unmasked points, which we denote by $\Delta_\lambda$.
Qualitatively, one expects this ratio to be of order of the
noise-to-signal ratio.  This expectation was confirmed by the
simulations, which yield $\Delta_\lambda = 10.9 \pm 4.8\%$ for $S/N =
10$ and $4.6 \pm 0.4\%$ for $S/N = 20$. In the data fits we typically
obtain values of 2--5\% for this figure of merit
(Table~\ref{tab:synthesis}), which is indeed of order of
$(S/N)^{-1}$.

The random uncertainties in the model parameters were estimated by
means of Monte Carlo simulations, adding Gaussian noise with amplitude
equal to the rms in the 4760--4810 \AA\ range and repeating the fits
100 times for each galaxy. The resulting dispersions in the individual
$x_j$ components range from 2 to 8\%, with an average $\sigma(x_j)$ of
4\%. As expected, the binned proportions are better determined than
the individual ones, with typical one sigma uncertainties of 2, 2, 4,
and 4\% for $x_{FC}$, $x_Y$, $x_I$ and $x_O$ respectively.  These
values should be regarded as order of magnitude estimates of the
errors in synthesis. The detailed mapping of the error domain in
parameter space requires a thorough investigation of the full
covariance matrix, a complex calculation given the high-dimensionality
of the problem.

\subsection{The diversity of stellar populations in Seyfert 2s}

\begin{figure*}
\resizebox{\textwidth}{!}{\includegraphics{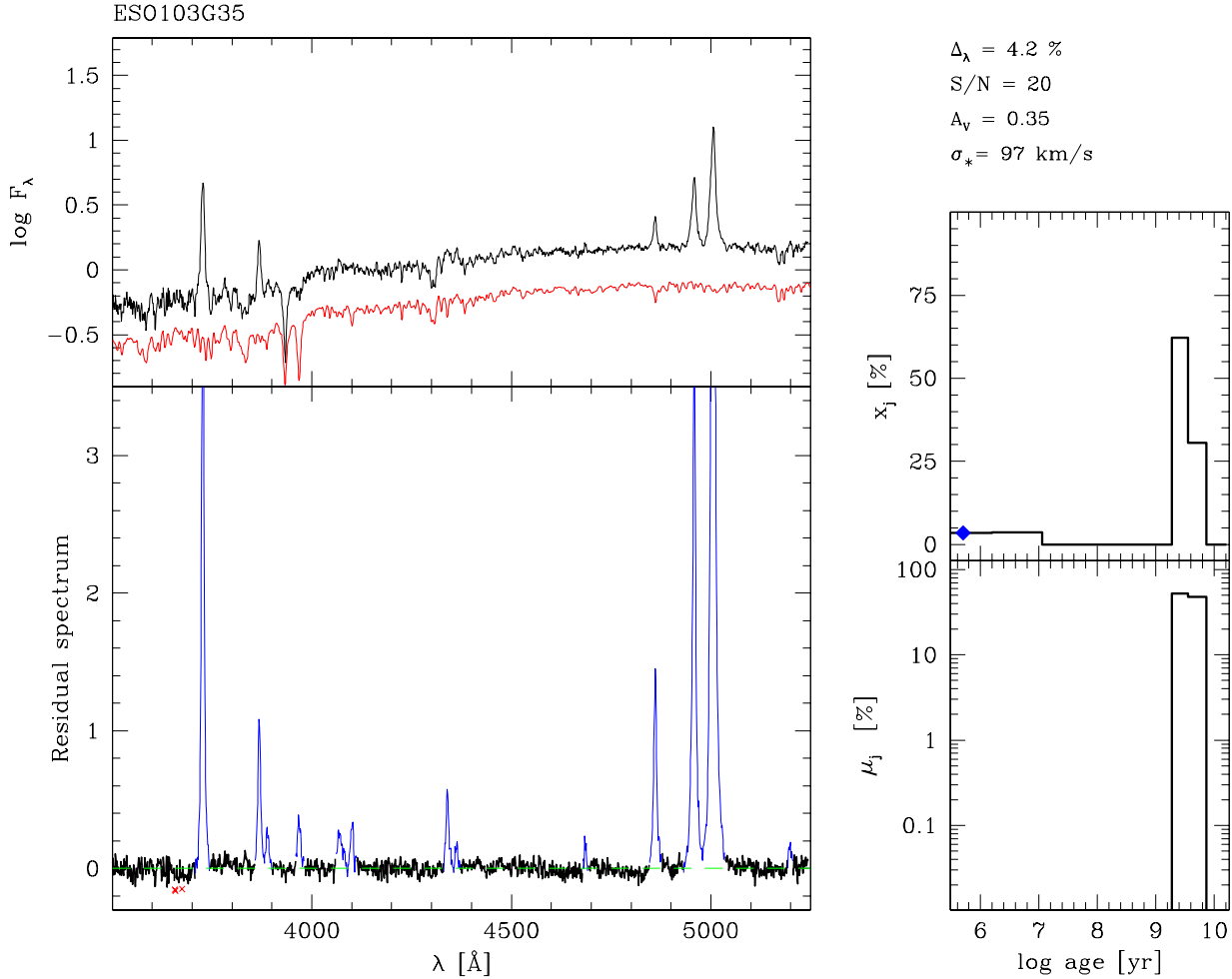}}
\caption{Results of the spectral fits for ESO 103-G35. The top left
panel shows the logarithm of the observed and the synthetic spectra
(shifted down for clarity). The $O_\lambda - M_\lambda$ residual
spectrum is shown in the bottom left panel.  Spectral regions actually
used in the synthesis are plotted with a thick line, while masked
regions are plotted with a thin line.  Crosses indicate points
excluded from the synthesis by a 3 sigma clipping algorithm. All
spectra are shown in units of the normalization flux $O_N$. Panels on
the right show the population vector binned in the 10 ages of the
base. The top panel corresponds to the population vector in flux
fraction ($\vec{x}$), normalized to $\lambda_0 = 4020$ \AA, while the
corresponding mass fractions vector $\vec{\mu}$ is shown in the
bottom. The power-law component $x_{FC}$ is plotted with an
(arbitrary) age of $10^{5.5}$ yr and marked by a diamond. No mass is
associated to this component (ie, $\mu_{FC} = 0$ in the bottom-right
panel).}
\label{fig:Fit_j001}
\end{figure*}

\begin{figure*}
\resizebox{\textwidth}{!}{\includegraphics{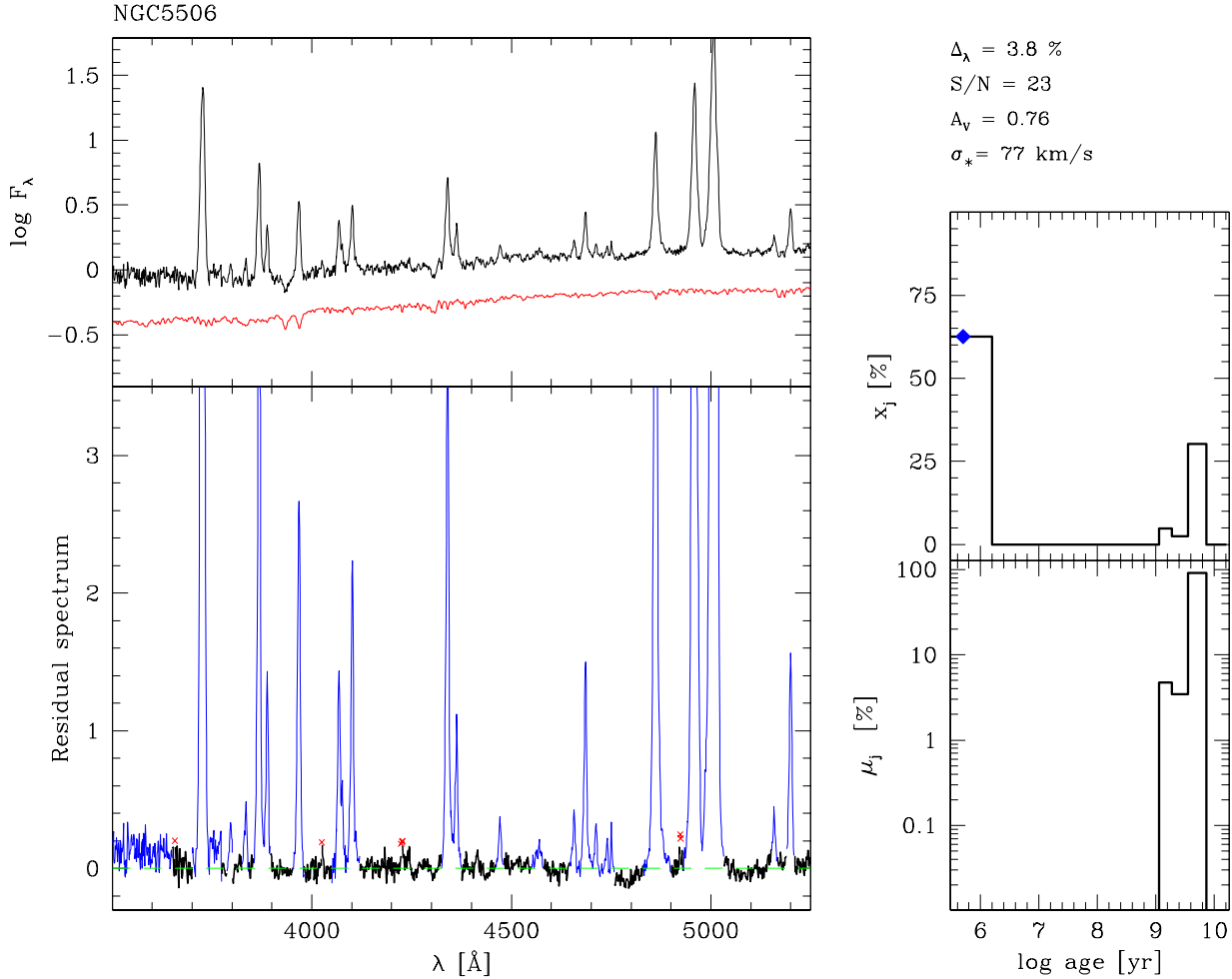}}
\caption{As figure \ref{fig:Fit_j001}, but for NGC 5506}
\label{fig:Fit_j062}
\end{figure*}

\begin{figure*}
\resizebox{\textwidth}{!}{\includegraphics{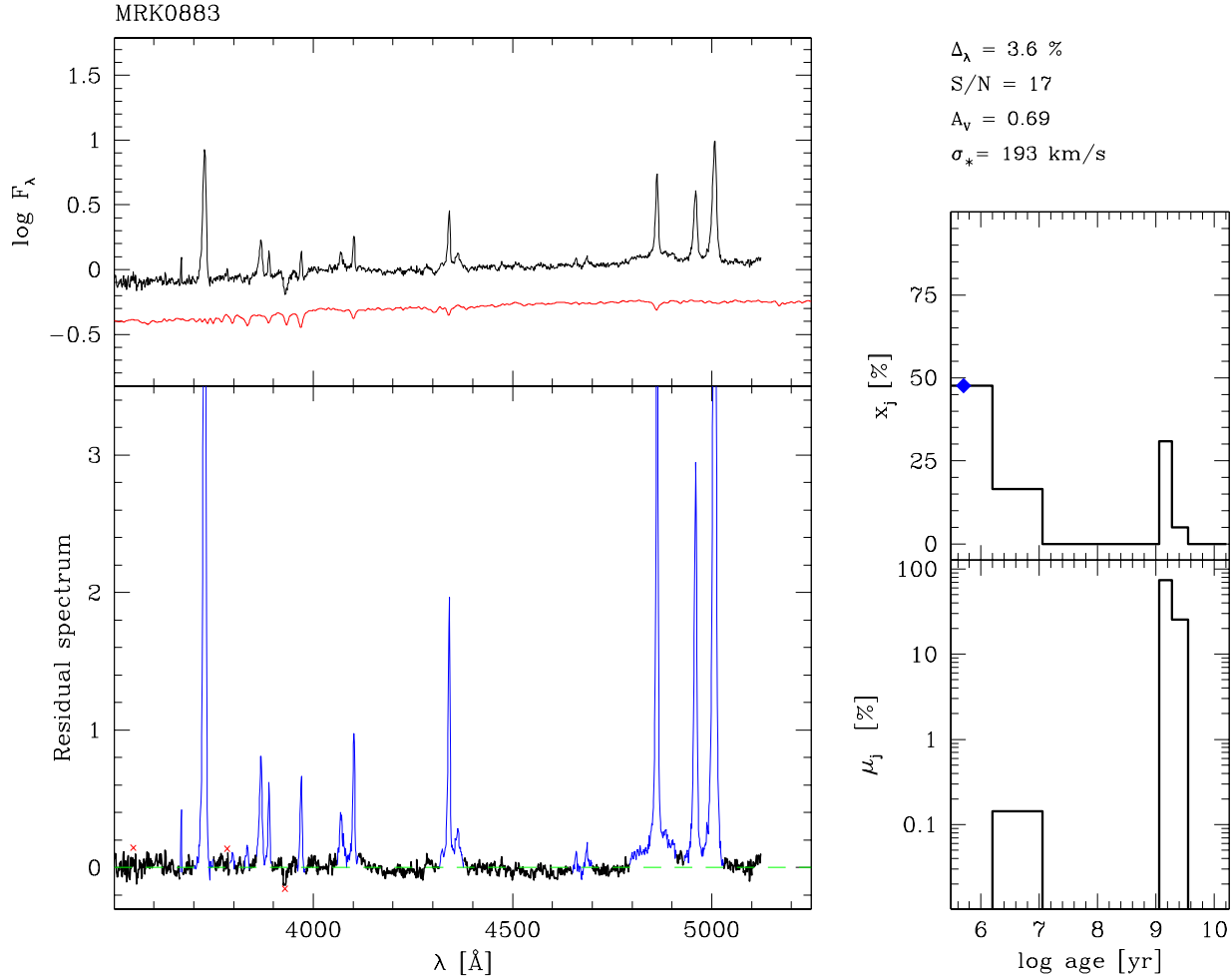}}
\caption{As figure \ref{fig:Fit_j001}, but for Mrk 883.}
\label{fig:Fit_j023}
\end{figure*}

\begin{figure*}
\resizebox{\textwidth}{!}{\includegraphics{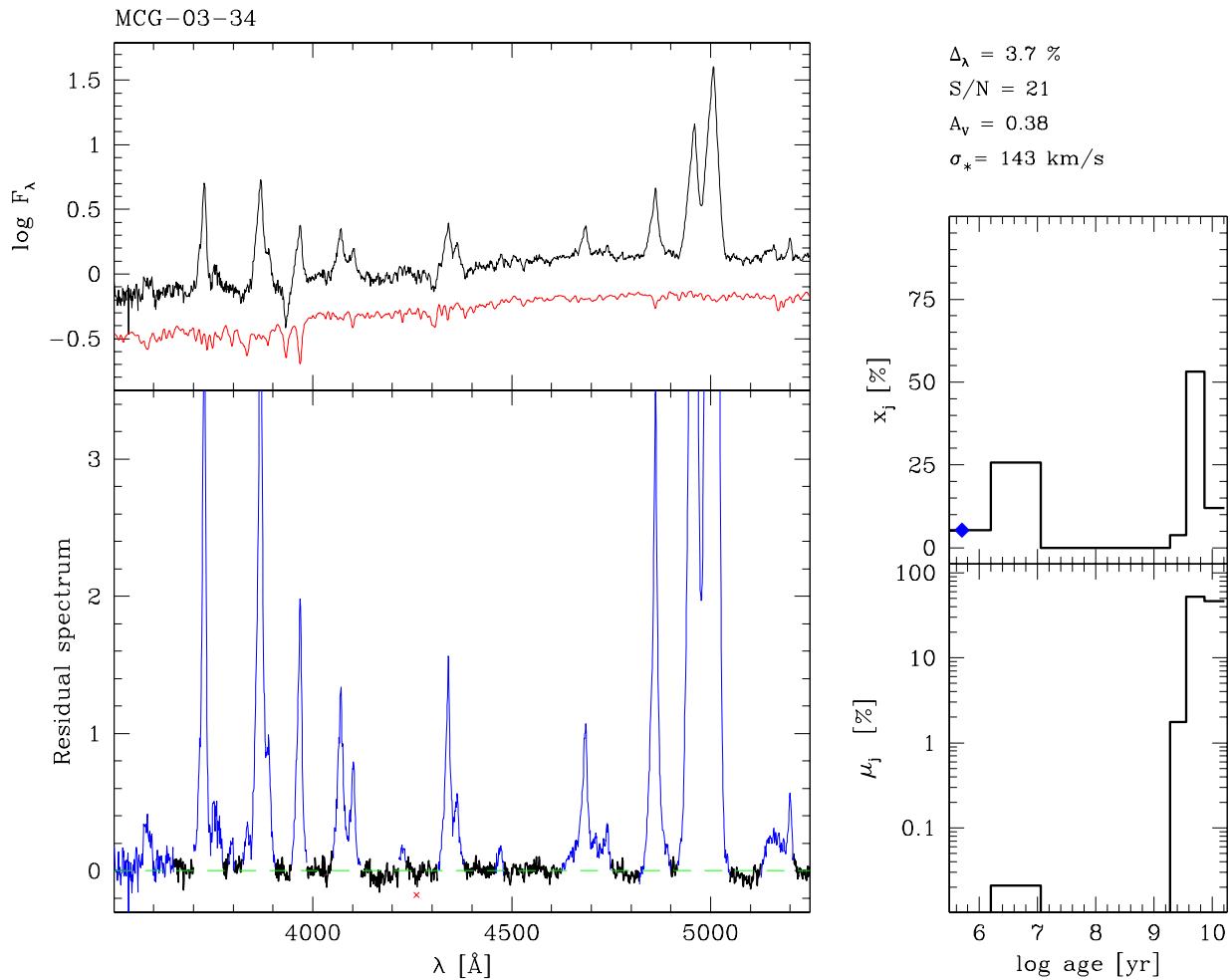}}
\caption{As figure \ref{fig:Fit_j001}, but for MGC -03-034.}
\label{fig:Fit_j022}
\end{figure*}

\begin{figure*}
\resizebox{\textwidth}{!}{\includegraphics{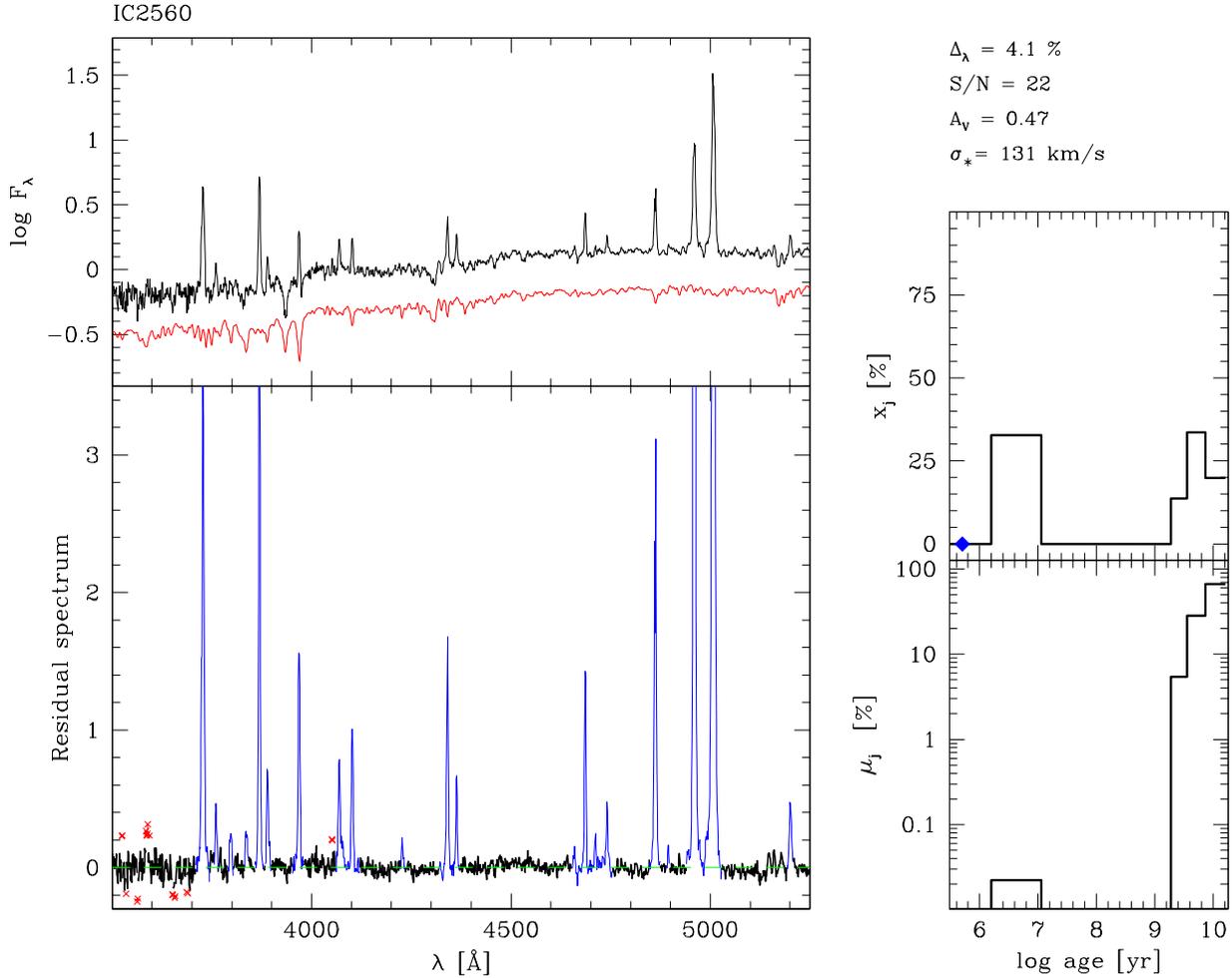}}
\caption{As figure \ref{fig:Fit_j001}, but for IC 2560.}
\label{fig:Fit_j017}
\end{figure*}

\begin{figure*}
\resizebox{\textwidth}{!}{\includegraphics{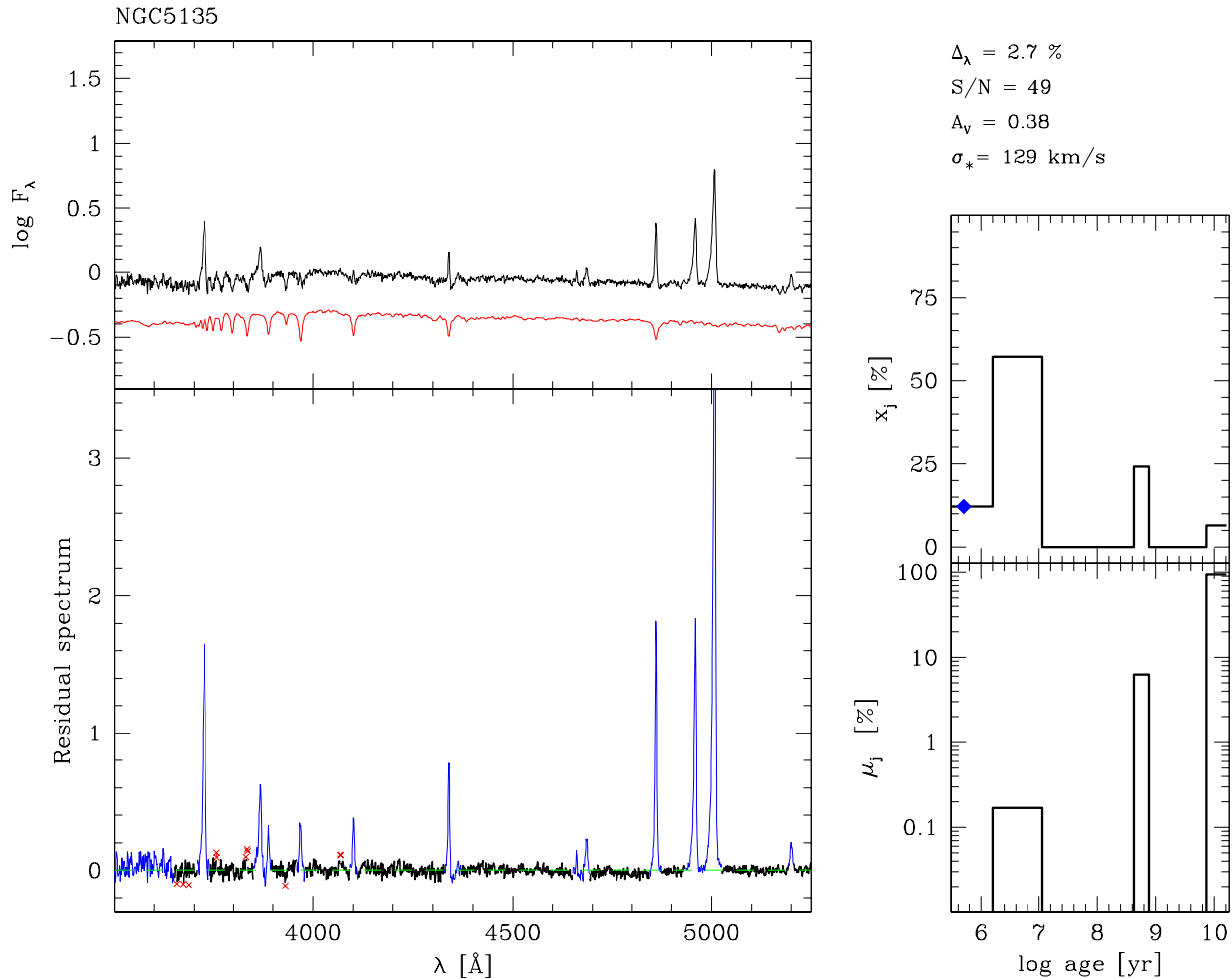}}
\caption{As figure \ref{fig:Fit_j001}, but for NGC 5135.}
\label{fig:Fit_j059}
\end{figure*}

\begin{figure*}
\resizebox{\textwidth}{!}{\includegraphics{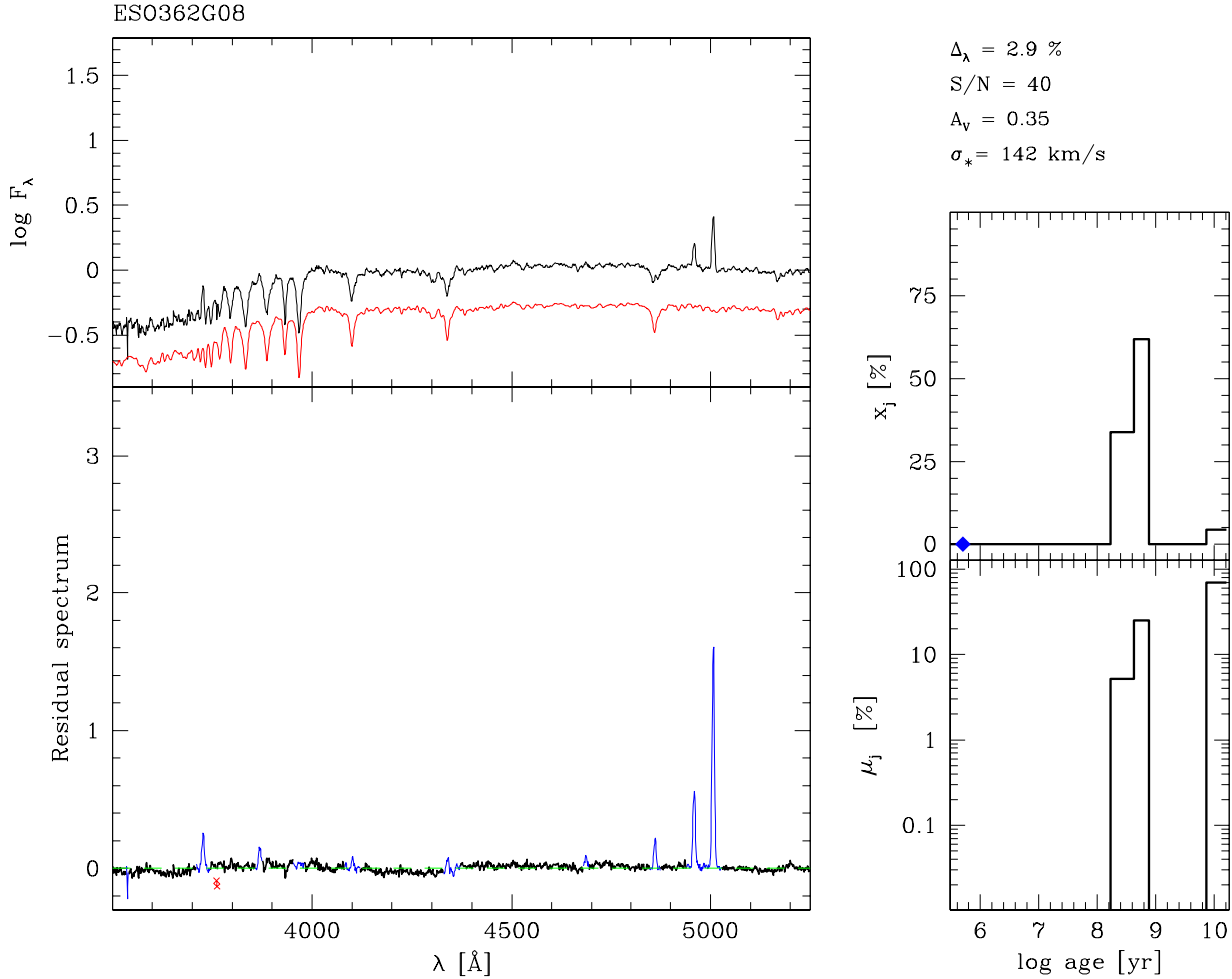}}
\caption{As figure \ref{fig:Fit_j001}, but for ESO 362-G08.}
\label{fig:Fit_j008}
\end{figure*}

\begin{figure*}
\resizebox{\textwidth}{!}{\includegraphics{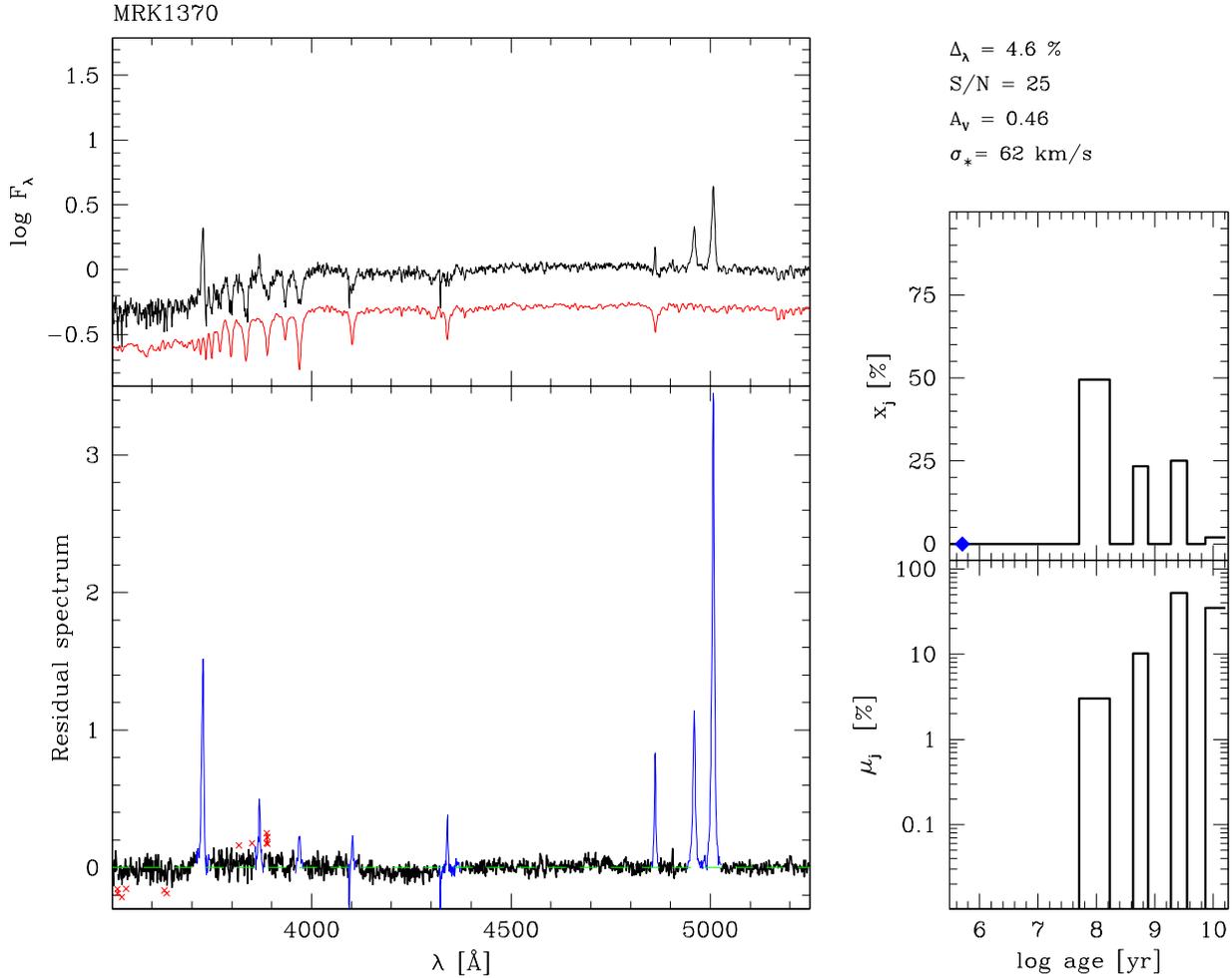}}
\caption{As figure \ref{fig:Fit_j001}, but for Mrk 1370.}
\label{fig:Fit_j026}
\end{figure*}

\begin{figure*}
\resizebox{\textwidth}{!}{\includegraphics{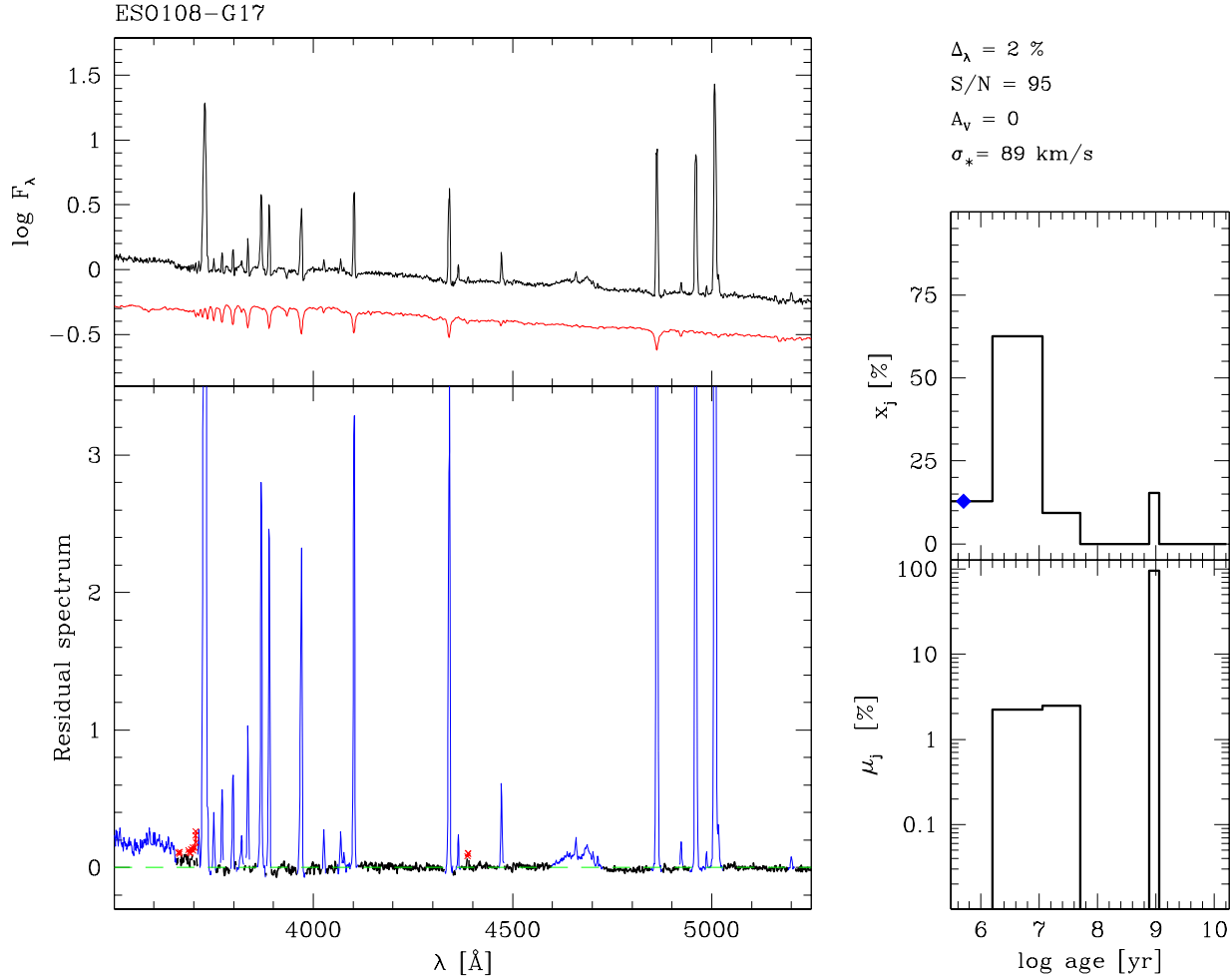}}
\caption{As figure \ref{fig:Fit_j001}, but for the WR galaxy ESO
108-G17.}
\label{fig:Fit_j003}
\end{figure*}

The spectral synthesis analysis shows that Seyfert 2 nuclei are
surrounded by virtually every type of stellar population, as can be seen
by the substantial variations in the derived SFHs (figures
\ref{fig:Fit_j001}--\ref{fig:Fit_j003} and Table \ref{tab:synthesis}).

Some nuclei, like ESO 103-G35 (Fig \ref{fig:Fit_j001}), NGC 4903, and
NGC 4941 are dominated by old stars, with populations older than 2 Gyr
accounting for over 80\% of the light at 4020 \AA, while ESO 362-G08
(Fig \ref{fig:Fit_j008}), Mrk 1370, NGC 2979 and others have very
strong intermediate age populations, with $x_I > 70\%$.  Young
starbursts are also ubiquitous. In Mrk 1210, NGC 5135 and NGC 7410
they account for more than half of the $\lambda$4020 flux, and in
several other Seyfert 2s their contribution exceeds 20\%.  Strong FC
components are also detected in several cases (eg, NGC 5506 and Mrk
883). In general, at least three of these four components are present
with significant strengths ($x \ga 10\%$) in any one galaxy.

\begin{figure}
  \includegraphics[width=8.8cm]{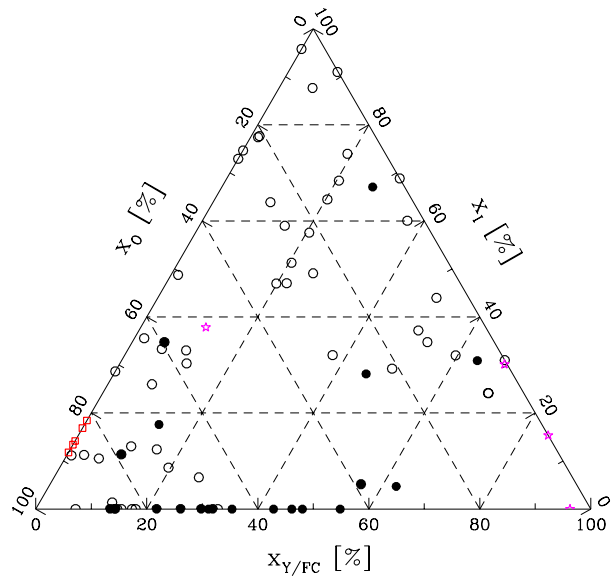}
\caption{Results of the spectral synthesis condensed on a
($x_{Y/FC},x_I,x_O$) description and plotted in a face on projection
of the $x_{Y/FC} + x_I + x_O = 1$ plane.  Squares denote normal
galaxies and the stars mark starburst galaxies. AGN (mostly Seyfert
2s) are plotted with circles. Filled circles indicate objects in which
broad lines have been identified either in direct or polarized
light.}
\label{fig:YIO_triangle}
\end{figure}

This diversity is further illustrated in Fig \ref{fig:YIO_triangle},
where we present a face-on projection of the $x_{Y/FC} + x_I + x_O =
1$ plane, with $x_{Y/FC} \equiv x_Y + x_{FC}$. This heterogeneity
contrasts with the results for our small sample of normal galaxies,
all five of which are dominated by old stars ($x_O \sim
85\%$), with a minor contribution from intermediate age populations.

\subsection{Comparison with previous results}

\label{sec:CompareWithCF01}

Several galaxies in our sample had their stellar populations studied
before. In this section we compare our results with those reported by
CF01, who combines observations of Seyfert 2s from several other
papers (Heckman \etal 1997; Storchi-Bergmann \etal 1998; Schmitt \etal
1999; Gonz\'alez Delgado \etal 1998; 2001). There are 13 galaxies in
common with CF01. Their data are similar in wavelength range and
quality to those in this paper. Furthermore (not coincidentally, of
course), they follow a description of stellar population plus FC
components based on three broad groups, similar to those defined in \S
\ref{sec:GroupingSchemes} and \ref{sec:J01Synthesis_Results}, which
facilitates comparison with our results. The main differences between
these two studies resides in the method to analyze the stellar
population mixture. Whereas we make use of the full spectrum, the
population synthesis performed by CF01 is based in a handful of
spectral indices (those listed in Table
\ref{tab:Stellar_Indices}). Also, the stellar population base used by
CF01 is that of Schmidt \etal (1991), built upon integrated spectra of
star clusters whose age and metallicity were determined by Bica \&
Alloin (1986), whereas here we use the BC03 models.  Notwithstanding
the different methodologies, as well as differences in the spectra
themselves (eg, due to the larger apertures in CF01), one expects to
find a reasonable agreement between these two studies.

\begin{figure}
\includegraphics[width=9cm]{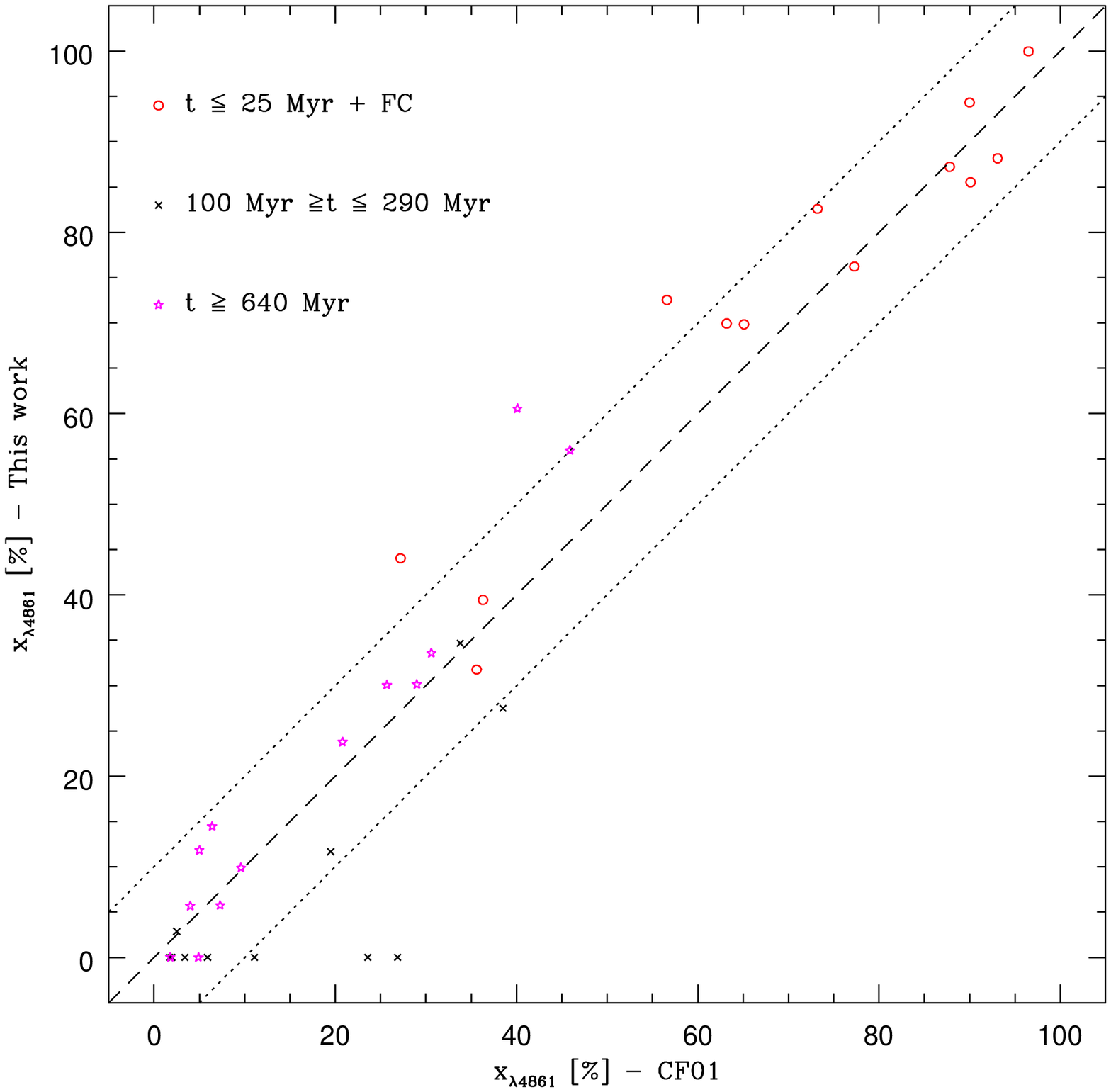}
\caption{Comparison of the population vector obtained in this work
with those in Cid Fernandes \etal (2001a) for 13 Seyfert 2s in
common. Different symbols indicate different age ranges. The identity
line is traced by the dashed line, while dotted lines indicate the
$\pm 10\%$ range.}
\label{fig:x_US_X_CF01}
\end{figure}

Figure \ref{fig:x_US_X_CF01} summarizes the results of this
comparison. Stars are used to plot the $x_{Y/FC}$ fraction of CF01
(called just ``$x_{FC}$'' in that paper) against our estimate, after
re-scaling it to $\lambda_0 = 4861$ \AA\ to match the normalization in
CF01. The agreement is very good, with a mean offset of 4\% and rms of
6\% between the two studies. Crosses and circles are used for the
$x_{INT}$ and $x_{OLD}$ components of CF01, which represent $\sim 100$
Myr and $\ga 1$ Gyr age bins. While qualitatively similar, these two
components do not compare well with our definitions of $x_I$ and
$x_O$, which are systematically larger and smaller than $x_{INT}$ and
$x_{OLD}$ respectively. This is essentially due to the different
age-grouping schemes employed. Binning our 100 and 290 Myr base
components into $x_I^\prime$ and all $t \ge 640$ Myr components into
$x_O^\prime$ yields a good correspondence with the $x_{INT}$ and
$x_{OLD}$ of CF01, with residuals similar to those obtained for
$x_{Y/FC}$, as shown in figure \ref{fig:x_US_X_CF01}.

The two discrepant crosses in the bottom of figure
\ref{fig:x_US_X_CF01} are IC 5135 (= NGC 7130) and NGC 5135. Again,
the disagreement is only apparent.  Our spectral synthesis find null
contributions of the 100 and 290 Myr components, but for NGC 5135 it
fits 27\% of the continuum at 4861 \AA\ as due to the 640 Myr
component, while for IC 5135 18 and 21\% are associated with 640 and
900 Myr-old bursts respectively.  In our discrete base these
components are adjacent in age to the 290 Myr one used as the upper
limit for $x_I^\prime$ in the plot, and hence could well be regrouped
within an intermediate age component by a slight modification of the
age range associated with $x_I^\prime$, removing the discrepancy. In
short, these ambiguities arise primarily because of the different age
resolutions employed in the two studies.  Finally, we note that our
results for these two galaxies compare well with those of Gonz\'alez
Delgado \etal (2001), who performed a detailed spectral fit of the
Balmer absorption lines characteristic of intermediate age populations
using their own models (Gonz\'alez Delgado, Leitherer \& Heckman
1999). For both IC 5135 and NGC 5135 they find $\sim 50$, 40 and 10\%
contributions (at 4800 \AA) of young, intermediate age and old
populations respectively, where the intermediate age component is
modeled as a combination of 200, 500 and 1000 Myr bursts. Our
condensed population vector (at 4861 \AA) for these galaxies are
$(x_{Y/FC},x_I,x_O) =$ (60,40,0) and (56,27,17) respectively.

We thus conclude that the population synthesis results reported here
are consistent to within $\Delta x \sim 10\%$ with previous results,
which is quite remarkable given differences in data and analysis
methods. A clear advantage of our approach with respect to methods
like those in CF01 is that we model the full spectrum, which, besides
providing more constraints, yields a more detailed understanding of
the different components which make up a Seyfert 2 spectrum. This
advantage is explored next.

\section{Analysis of the Starlight-subtracted spectra}

\label{sec:Starlight_Subtracted_Spectra}

The excellent fits of the starlight in our spectra provide a unique
opportunity to investigate the presence of weak emission features in
the $O_\lambda - M_\lambda$ residual spectra. In this section we
explore this opportunity and examine the consequences for the
interpretation of the synthesis results.

\subsection{Wolf Rayet stars}

\label{sec:WRbump}

The presence of WR starts, which produce the so called WR bump at
$\sim 4680$ \AA, provides a clear indication of starburst activity
which is complementary to the population synthesis.  We have used the
starlight subtracted spectra to investigate the presence of the WR
bump in our sample galaxies.

WR stars were detected in the Starburst/WR galaxy ESO108-G17 and in
the Seyfert 2s MCG-03-34, Mrk 1210 and NGC 424. Hints of a WR bump are
also present in ESO 138-G01 and NGC 4507, but these cases require
confirmation either from deeper observations or detection of the 5812
\AA\ feature (Schaerer, Contini \& Kunth 1999). These results are
presented as notes in Table~\ref{tab:synthesis}, which summarizes the
results of the spectral synthesis models.  Galaxies that show WR
signatures present important young stellar components (average $x_Y$
of 30\%), thus validating the results of the synthesis.  Given the
severe blending with nebular lines of HeII, ArIV, NeIV and FeIII in
Seyfert 2s, it is probably futile to attempt any detailed modeling of
the WR population.

\subsection{Broad emission lines}

\label{sec:data_BroadLines}

\begin{figure}
\includegraphics[width=18cm]{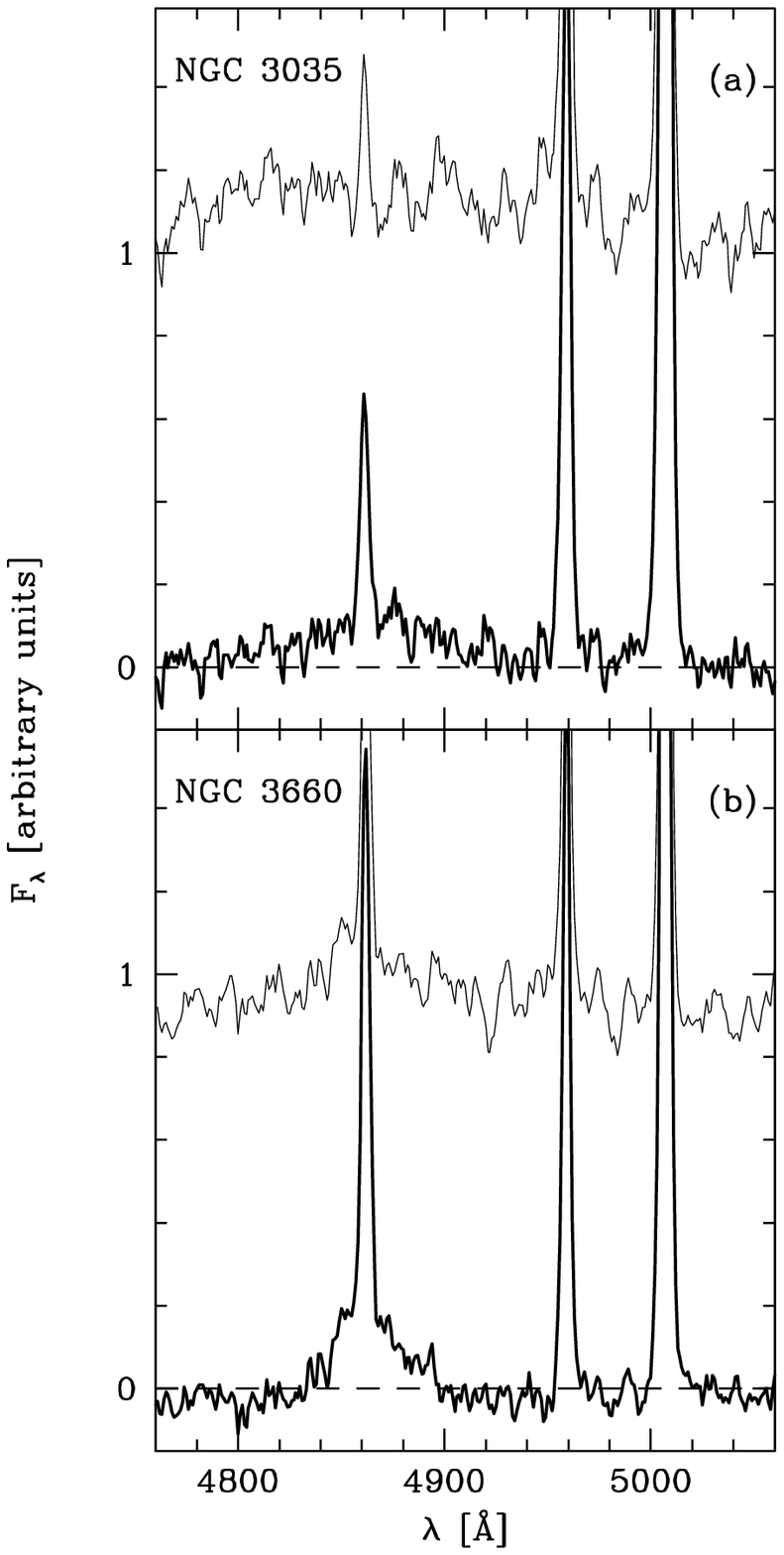}
\caption{(a) Top: Total spectrum of NGC 3035. Bottom: Starlight
subtracted spectrum of the same galaxy, showing a broad H$\beta$
component not evident in the total spectrum. The dashed line marks the
zero flux level for the bottom spectrum. (b) As the top panel, but for
NGC 3660.}
\label{fig:NGC3035_and_NGC3660}
\end{figure}

A close inspection of the starlight subtracted spectra reveals a weak
broad H$\beta$ in several of our Seyfert 2s. This is the case of NGC
3035, classified as a Seyfert 1 in NED but reclassified as a Seyfert 2
by J01 on the basis of the absence of broad lines. The top spectrum in
figure \ref{fig:NGC3035_and_NGC3660}a is the same one used by
J01. When we subtract our starlight model, however, a weak but clear
broad component emerges in H$\beta$ (bottom spectrum in figure
\ref{fig:NGC3035_and_NGC3660}a).

At least seven other objects exhibit a similar weak broad component
under H$\beta$: ESO 383-G18, MCG-03-34-064, NGC 424, NGC 1068, NGC
3660 (figure \ref{fig:NGC3035_and_NGC3660}b), NGC 5506 and NGC 7212,
and 2 other objects present marginal evidence for a broad H$\beta$.
In principle, these nuclei could be reclassified as Seyfert 1.5 or
1.8, following the notation of Osterbrock (1984). However, the
detection of weak broad lines is not {\it per se} reason enough to
classify a nucleus as of type 1. This violation of the historical
definition of Seyfert types by Kachikian \& Weedman (1974) is allowed
in the context of the unified model, which predicts that a weak BLR
should be detected in Seyfert 2s (at least those with a scattering
region). We have thus opted not to reclassify these objects, with the
understanding that these weak features are probably scattered lines.

The detection of a BLR provides a useful constraint for the
interpretation of the spectral synthesis results. Broad emission lines
in AGN always come with an associated non-stellar FC, presumably from
the nuclear accretion disk. This is true regardless of whether the
nucleus and BLR are seen directly, as in {\it bona fide} Seyfert 1s,
or via scattering, as in NGC 1068, the prototypical Seyfert 2
(Antonucci \& Miller 1985). Hence, when the BLR is detected we can be
sure that an FC component is also present. 

Column 13 of Table \ref{tab:synthesis} lists the spectropolarimetry
information available for galaxies in our sample (from the compilation
by Gu \& Huang 2002 plus new data from Lumsden, Alexander \& Hough
2004). The existence of an AGN FC can also be established {\it a
priori} in Seyfert 2s where spectropolarimetry reveals a type 1
spectrum, ie, those with a ``Hidden BLR''. This is the case of 17
objects in our sample. In 7 of these we were able to identify a weak
broad H$\beta$ in our direct spectra, while in the remaining cases the
BLR detected with the aid of polarimetry is just too faint to be
discerned in our data. In these cases, we expect a correspondingly
weak FC component.  Spectropolarimetry data are available for other
fourteen galaxies in our list, but no hidden Seyfert 1 was detected,
either because of the absence of an effective mirror or observational
limitations.  A further alternative is that these are ``genuine
Seyfert 2s'', ie, AGN with no BLR (Tran 2001, but see Lumsden \etal
2004).

Intriguingly, Tran (2001) reports no BLR detection from his
spectropolarimetry data on NGC 3660, while figure
\ref{fig:NGC3035_and_NGC3660}b leaves no doubt as to the existence of
this component in direct light. Variability could be the cause for
this discrepancy, in which case this galaxy should be classified as a
type 1 Seyfert, because the BLR is seen directly.  A type transition
of this sort has happened in NGC 7582 (Aretxaga \etal 1999), which
suddenly changed spectrum from that of a Seyfert 2 to one with broad
emission lines sometime in mid 1998.

Table \ref{tab:synthesis} lists the galaxies for which we have
evidence for the presence of weak broad emission lines, either from
spectropolarimetry data in the literature (HBLR) or from the present
study (BLR). There is a clear tendency for these ``broad line Seyfert
2s'' (BLS2s) to also show an important FC component in the synthetic
spectra.  We thus conclude that the synthesis is able to identify a
{\it bona fide} non-stellar continuum when one is present.  Further
confirmation of this fact comes from the synthesis of the Seyfert 1
nucleus Mrk 883, which yields $x_{FC} = 48\%$.  As shown in Paper II
(Gu \etal, in preparation), there is also a tendency for BLS2s to have
larger [OIII] luminosities. This is consistent with them having more
powerful AGN, which, for a fixed scattering efficiency, implies a
brighter, easier to detect scattered FC, in agreement with the results
reported above.

\subsection{FC versus Young Stars}

\label{sec:FC_X_YoungStars}

The synthesis also reveals a strong FC component in several objects
for which there is no evidence of a BLR. The question then arises of
whether $x_{FC}$ corresponds truly to an AGN continuum or to a dusty
starburst.

It is hard to distinguish a $F_\nu \propto \nu^{-1.5}$ power-law from
the spectrum of a reddened young starburst over the wavelength range
spanned by our data.  The main differences between the spectrum of a 5
Myr SSP seen through $A_V \sim 2$--3 mag of dust and this power-law
lie in the Balmer absorption lines and in the blue side of the Balmer
jump. Since these features are masked in the synthesis, the $x_{FC}$
strength can be attributed to either a dusty-starburst, a true AGN FC,
or to a combination of these. This confusion is likely to occur in our
synthesis models given that we impose a common $A_V$ for all base
components while young starbursts are intrinsically rather dusty
(Selman \etal 1999).

To illustrate this issue, we note that we obtain $x_{FC} = 13\%$ for
the WR galaxy ESO 108-G17 (Fig \ref{fig:Fit_j003}). Given that there
is no indication of an AGN in this irregular galaxy, it is very likely
that this FC component is actually associated with a dusty starburst.
The same applies to the starbursting merger NGC 3256, for which we
derive $x_{FC} = 24\%$. Note, however, that in both cases the
synthesis identifies a dominant young population, with $x_Y = 72$ and
45\% respectively.

Not surprisingly, the starburst-FC degeneracy is more pronounced in
galaxies known to harbour an active nucleus.  For NGC 6221, for
instance, we find $x_{FC} = 32\%$, $x_Y = 28\%$ and $A_V = 0.5$, while
from the detailed imaging analysis and modeling of the full spectral
energy distribution carried out by Levenson \etal (2001) we know that
this nucleus is dominated by heavily extincted ($A_V \sim 3$) young
star clusters within $< 1^{\prime\prime}$ of the AGN, with just a few
percent of the optical light originating in the AGN responsible for
its Seyfert-like X-ray properties.  Most of the large $x_{FC}$
obtained by the synthesis is thus associated with this reddened
starburst.  Another emblematic example is IC 5135 (NGC 7130), whose
UV spectrum shows wind lines of massive stars and a logarithmic slope
typical of dusty starbursts, with little or no sign of a power-law
(Gonz\'alez Delgado \etal 1998), while our synthesis yields $x_{FC} =
35\%$, $x_Y = 34\%$ and $A_V = 0$.  Similar comments can be made about
other galaxies in the sample (eg, NGC 5135 and NGC 7582).

These examples show that dusty starbursts, when present, are detected
as an FC component by our synthesis method. The fraction $x_{FC}$
should thus be regarded as an {\it upper limit} to the contribution of
a true FC, while the light fraction associated with very recent
star-formation is bracketed between $x_Y$ and $x_{Y/FC}$.  While there
is no way to break this spectral degeneracy without broader spectral
coverage, the information on the presence of weak broad lines,
collected in Table \ref{tab:synthesis}, helps tackling this issue. As
discussed in \S\ref{sec:data_BroadLines}, whenever a BLR component is
detected either in direct or polarized light we can be sure that a
true FC is present. One thus expects BLS2s to have larger $x_{FC}$
values than Seyfert 2s for which we have no such independent evidence
for the presence of an AGN continuum.

\begin{figure}
\includegraphics[width=9cm]{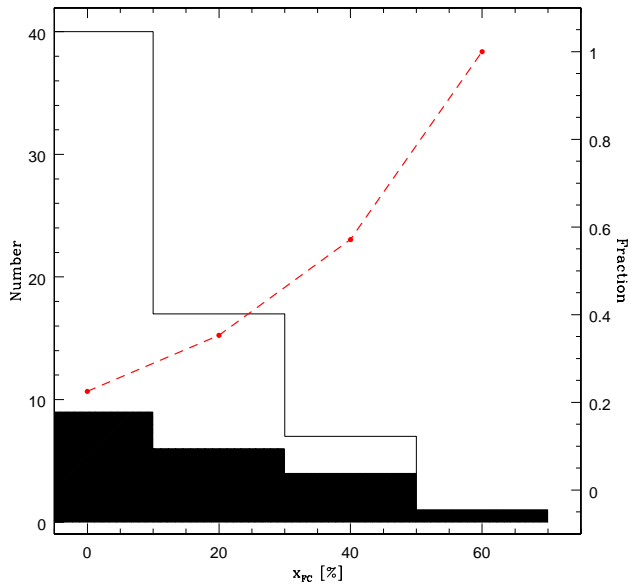}
\caption{Histogram of the FC fraction in Seyfert 2s. Filled areas
correspond to galaxies where a true FC is known to exist from the
detection of a BLR in either direct or polarized light.  The fraction
of these latter galaxies with respect to the whole sample increases
with increasing $x_{FC}$, as shown by the dashed line and right-hand
scale.}
\label{fig:xFC_histogram}
\end{figure}

This expectation is confirmed by the synthesis. The four largest
values of $x_{FC}$ among Seyfert 2s, for instance, are all found in
systems which show weak broad H$\beta$. The link between large
$x_{FC}$ and the presence of broad lines is illustrated in figure
\ref{fig:xFC_histogram}, which shows that the fraction of BLS2s
increases systematically with increasing $x_{FC}$. This effect is also
clear in the statistics of $x_{FC}$, which assumes a median (mean)
value of 21\% (19\%) among BLS2s but just 2\% (7\%) among the rest of
the Seyfert 2s. As predicted in \S\ref{sec:data_BroadLines}, selecting
only the galaxies where we see the BLR in our direct spectra increases
the median (mean) $x_{FC}$ to 31\% (29\%), while the ten nuclei where
a hidden BLR appears in polarized spectra but not in our data have
weaker FCs: 5\% (8\%).

Overall, these results confirm the prediction of Cid Fernandes \&
Terlevich (1995), who estimated that a broad component in H$\beta$
should become discernible whenever the scattered FC contributes with
$\ga 20$\% of the optical continuum. Whenever $x_{FC}$ exceeds $\sim
20\%$ but no BLR is seen, this component most likely originates in a
dusty starburst rather than in a non-stellar source.

\section{Stellar Indices}

\label{sec:Indices}

In order to facilitate the comparison of some relevant aspects of our
population synthesis models and published work, we have found it
convenient to express our results in terms of specific indices (lines
and colours).  In this section we present a simple characterization of
the stellar populations in Seyfert 2s on the basis of a set of indices
commonly used in the literature.

\subsection{Direct measurements}

\label{sec:Direct_Indices}

In order to provide an empirical characterization of stellar
populations in our sample galaxies, we have measured a set of stellar
indices directly from the observed spectra.  To facilitate the
comparison with previous studies, we have employed the index
definitions used by Cid Fernandes \etal (2004), which are ultimately
based on the studies by Bica \& Alloin (1986a,b) and Bica (1988) of
star cluster and galaxy spectra.

We have measured the equivalent widths of the Ca II K ($W_K$), CN
($W_{CN}$) and G bands ($W_G$), as well as the colours $C_{3660}$ and
$C_{4510}$, defined as ratios between the continuum at 3660 and 4510
\AA\ to the continuum at 4020 \AA, respectively. All these indices are
measured with respect to a pseudo continuum, which we define
automatically following the recipe in Cid Fernandes \etal (2004),
after verifying that it works well for the present sample.  Table
\ref{tab:Stellar_Indices} lists the resulting stellar indices.  In
this paper we give particular emphasis to the K line. Previous studies
have shown that $W_K$, which essentially measures the contrast between
old and young populations, provides a useful mono-parametric
description of stellar populations in galaxies. For instance, Seyfert
2s with unambiguous signatures of recent star-formation (such as UV
wind lines, the WR bump, or high order Balmer absorption lines) all
have K lines diluted to $W_K \la 10$ \AA\ (CF01).

\begin{figure}
\includegraphics[width=18cm]{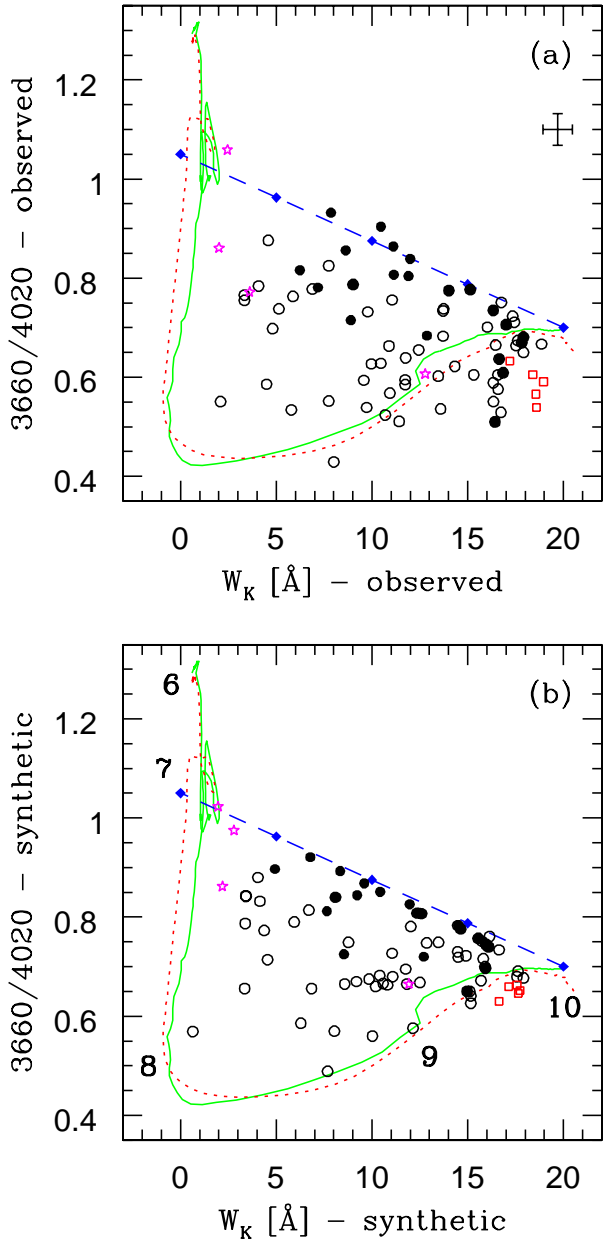}
\caption{$C_{3660}$ versus $W_K$ diagram for the J01 sample. Panel (a)
shows the results for $C_{3660}$ and $W_K$ as measured directly from
the observed spectra, whereas in (b) these two indices are measured
from synthetic spectra constructed to match the observed stellar
spectrum. In both panels the solid (green) and dotted (red) curves
represent the evolution of BC03 models of instantaneous bursts with
$Z_\odot$ and $2.5 Z_\odot$ metallicity respectively. The approximate
location of populations with ages $\log t = 6$, 7 , 8 , 9 and 10 yr
are indicated by the labels in panel (b). The dashed (blue) line
between $(W_K,C_{3660}) = (0,1.05)$ and $\sim (20,0.7)$ represents a
mixing line, where a $F_\nu \propto \nu^{-1.5}$ power-law is added to
a $10^{10}$ yr stellar population; diamonds mark power-law fractions
in steps of 25\%. The average error bar is indicated in the top-right
in (a). Symbols as in figure \ref{fig:YIO_triangle}.}
\label{fig:C3660_x_WK}
\end{figure}

Figure \ref{fig:C3660_x_WK}a shows the $W_K$ versus $C_{3660}$
diagram.  This plot has the advantage of involving only direct
measurements, which, in the case of the JO1 spectra, is not possible
for more popular indices.  The solid and dotted curves show the
evolution of instantaneous burst BC03 models for solar (solid, green
line) and 2.5 solar metallicity (dotted, red).  The region occupied by
pure stellar populations in this diagram is delimited by these tracks
and a mixing line joining the youngest and oldest models. Points
located below the BC03 models in the bottom-right of the plot are
explained as a combination of observational errors and intrinsic
reddening, both of which affect more the $C_{3660}$ colour than $W_K$.

The dashed line in figure \ref{fig:C3660_x_WK} represents a mixing
line, where a canonical AGN $F_\nu \propto \nu^{-1.5}$ power-law is
added to an old stellar population in different proportions, from a
pure power-law at $(W_K,C_{3660}) = (0,1.05)$ to a pure $10^{10}$ yr
population at $\sim (20,0.7)$. The fact that few galaxies line up
along this mixing line shows that a simple ``elliptical galaxy'' plus
power-law model, adopted in many early AGN studies, does not apply to
the bulk of Seyfert 2s.  On the contrary, the spread of points in this
diagram implies a substantial diversity in the central stellar
populations of Seyfert 2s, in agreement with our synthesis results
(eg, figure \ref{fig:YIO_triangle}) as well as with previous studies
(Cid Fernandes, Storchi-Bergmann \& Schmitt 1998; J01; Serote Roos
\etal 1998; Boisson \etal 2000).

\begin{table}
\tiny
\begin{centering}
\begin{tabular}{lrrrrr}
\multicolumn{6}{c}{Stellar Indices}\\ \hline
Galaxy          &
$W_K$ [\AA]     & 
$W_{CN}$ [\AA]  & 
$W_G$ [\AA]     & 
$C_{3660}$      &
$C_{4510}$      \\ \hline
ESO 103G35   &  16.6$\pm$0.7 &  10.5$\pm$0.9 &  10.1$\pm$0.4 & 0.61$\pm$0.03 & 1.47$\pm$0.04 \\
ESO 104G11   &  10.9$\pm$1.3 &   5.4$\pm$1.7 &   7.8$\pm$0.7 & 0.57$\pm$0.05 & 1.21$\pm$0.06 \\
ESO 137-G34  &  16.8$\pm$0.7 &  11.4$\pm$0.9 &  10.8$\pm$0.4 & 0.75$\pm$0.04 & 1.45$\pm$0.04 \\
ESO 138-G01  &   7.7$\pm$0.5 &   5.0$\pm$0.7 &   4.6$\pm$0.4 & 0.83$\pm$0.02 & 0.93$\pm$0.02 \\
ESO 269-G12  &  17.5$\pm$0.9 &  14.0$\pm$1.0 &  11.1$\pm$0.5 & 0.66$\pm$0.04 & 1.42$\pm$0.04 \\
ESO 323G32   &  16.3$\pm$0.8 &  13.8$\pm$0.9 &  10.7$\pm$0.4 & 0.55$\pm$0.03 & 1.39$\pm$0.04 \\
ESO 362G08   &   8.0$\pm$0.5 &   8.0$\pm$0.6 &   7.4$\pm$0.3 & 0.43$\pm$0.02 & 1.10$\pm$0.02 \\
ESO 373G29   &   5.9$\pm$0.5 &   6.4$\pm$0.7 &   6.4$\pm$0.4 & 0.76$\pm$0.02 & 1.02$\pm$0.02 \\
ESO 381-G08  &   8.9$\pm$0.4 &   5.5$\pm$0.5 &   5.6$\pm$0.2 & 0.72$\pm$0.01 & 1.15$\pm$0.02 \\
ESO 383-G18  &  10.5$\pm$0.8 &   6.1$\pm$1.1 &   7.0$\pm$0.6 & 0.90$\pm$0.04 & 1.07$\pm$0.04 \\
ESO 428G14   &  13.5$\pm$0.7 &  10.7$\pm$0.8 &   9.2$\pm$0.4 & 0.60$\pm$0.03 & 1.24$\pm$0.03 \\
ESO 434G40   &  16.7$\pm$0.7 &  13.5$\pm$0.9 &  11.0$\pm$0.4 & 0.64$\pm$0.03 & 1.52$\pm$0.04 \\
Fai 0334     &   9.8$\pm$1.8 &   7.5$\pm$1.9 &   8.6$\pm$0.9 & 0.73$\pm$0.07 & 1.19$\pm$0.06 \\
Fai 0341     &  13.7$\pm$1.2 &  10.0$\pm$1.3 &  10.5$\pm$0.6 & 0.73$\pm$0.05 & 1.49$\pm$0.06 \\
IC 1657      &  10.9$\pm$1.4 &   7.4$\pm$1.7 &   8.9$\pm$0.8 & 0.66$\pm$0.06 & 1.31$\pm$0.07 \\
IC 2560      &  11.0$\pm$1.0 &  11.0$\pm$1.1 &   9.6$\pm$0.5 & 0.76$\pm$0.04 & 1.40$\pm$0.05 \\
IC 5063      &  16.9$\pm$0.7 &  12.4$\pm$0.8 &  10.5$\pm$0.4 & 0.61$\pm$0.04 & 1.47$\pm$0.04 \\
IC 5135      &   4.6$\pm$0.4 &   3.4$\pm$0.5 &   3.2$\pm$0.3 & 0.88$\pm$0.02 & 0.92$\pm$0.01 \\
IRAS 11215   &  14.3$\pm$0.7 &   9.0$\pm$1.0 &   9.1$\pm$0.4 & 0.62$\pm$0.03 & 1.26$\pm$0.04 \\
MCG +01-27   &   7.8$\pm$0.7 &   8.2$\pm$0.8 &   7.4$\pm$0.4 & 0.55$\pm$0.03 & 1.05$\pm$0.03 \\
MCG -03-34   &  11.9$\pm$1.5 &  12.3$\pm$1.7 &   9.1$\pm$0.9 & 0.80$\pm$0.06 & 1.35$\pm$0.07 \\
MRK 0897     &   4.8$\pm$0.4 &   4.1$\pm$0.5 &   4.1$\pm$0.3 & 0.70$\pm$0.01 & 0.93$\pm$0.01 \\
MRK 1210     &   9.0$\pm$1.5 &   8.7$\pm$1.9 &   6.5$\pm$1.0 & 0.79$\pm$0.06 & 1.18$\pm$0.07 \\
MRK 1370     &   5.8$\pm$0.8 &   7.2$\pm$0.9 &   6.1$\pm$0.4 & 0.53$\pm$0.02 & 1.05$\pm$0.03 \\
NGC 0424     &  11.1$\pm$0.6 &   9.5$\pm$0.7 &   7.8$\pm$0.4 & 0.86$\pm$0.03 & 1.33$\pm$0.03 \\
NGC 0788     &  17.9$\pm$0.8 &  13.2$\pm$1.0 &  11.2$\pm$0.5 & 0.68$\pm$0.03 & 1.37$\pm$0.04 \\
NGC 1068     &   8.6$\pm$0.5 &   8.9$\pm$0.7 &   6.5$\pm$0.4 & 0.86$\pm$0.02 & 1.14$\pm$0.02 \\
NGC 1125     &  10.5$\pm$1.0 &   8.4$\pm$1.2 &   7.9$\pm$0.5 & 0.63$\pm$0.03 & 1.14$\pm$0.04 \\
NGC 1667     &  18.9$\pm$1.0 &  15.3$\pm$1.2 &  11.7$\pm$0.6 & 0.67$\pm$0.04 & 1.46$\pm$0.06 \\
NGC 1672     &  10.0$\pm$0.6 &   7.9$\pm$0.8 &   6.8$\pm$0.4 & 0.63$\pm$0.02 & 1.08$\pm$0.02 \\
NGC 2110     &  13.7$\pm$1.0 &  14.2$\pm$1.1 &  10.5$\pm$0.5 & 0.74$\pm$0.05 & 1.40$\pm$0.05 \\
NGC 2979     &  11.4$\pm$0.9 &   8.5$\pm$1.1 &   8.7$\pm$0.5 & 0.51$\pm$0.03 & 1.29$\pm$0.04 \\
NGC 2992     &  16.4$\pm$0.8 &  14.2$\pm$0.8 &  11.2$\pm$0.4 & 0.51$\pm$0.04 & 1.57$\pm$0.04 \\
NGC 3035     &  12.9$\pm$0.6 &  11.5$\pm$0.7 &   9.1$\pm$0.3 & 0.68$\pm$0.03 & 1.32$\pm$0.03 \\
NGC 3081     &  15.1$\pm$0.9 &  13.6$\pm$1.0 &   9.4$\pm$0.5 & 0.78$\pm$0.04 & 1.44$\pm$0.05 \\
NGC 3281     &  16.0$\pm$1.1 &  12.5$\pm$1.3 &  11.6$\pm$0.6 & 0.70$\pm$0.05 & 1.73$\pm$0.07 \\
NGC 3362     &  15.3$\pm$0.8 &  11.3$\pm$0.9 &   9.5$\pm$0.5 & 0.61$\pm$0.03 & 1.39$\pm$0.04 \\
NGC 3393     &  17.4$\pm$0.7 &  14.6$\pm$0.8 &  10.6$\pm$0.4 & 0.71$\pm$0.03 & 1.44$\pm$0.04 \\
NGC 3660     &   7.2$\pm$0.7 &   8.0$\pm$0.8 &   6.4$\pm$0.4 & 0.78$\pm$0.03 & 1.26$\pm$0.03 \\
NGC 4388     &  14.0$\pm$1.0 &   9.3$\pm$1.2 &   9.4$\pm$0.5 & 0.77$\pm$0.04 & 1.43$\pm$0.05 \\
NGC 4507     &  11.1$\pm$0.4 &   8.8$\pm$0.6 &   7.4$\pm$0.3 & 0.81$\pm$0.04 & 1.20$\pm$0.02 \\
NGC 4903     &  17.6$\pm$0.9 &  15.1$\pm$1.0 &  12.0$\pm$0.4 & 0.67$\pm$0.04 & 1.71$\pm$0.06 \\
NGC 4939     &  17.3$\pm$0.9 &  14.3$\pm$1.0 &  11.8$\pm$0.5 & 0.72$\pm$0.04 & 1.54$\pm$0.05 \\
NGC 4941     &  17.9$\pm$0.8 &  15.3$\pm$1.0 &  11.3$\pm$0.5 & 0.65$\pm$0.03 & 1.58$\pm$0.05 \\
NGC 4968     &  10.7$\pm$1.0 &   7.8$\pm$1.2 &   8.1$\pm$0.5 & 0.52$\pm$0.04 & 1.29$\pm$0.05 \\
NGC 5135     &   3.3$\pm$0.4 &   3.7$\pm$0.5 &   3.8$\pm$0.3 & 0.77$\pm$0.02 & 0.89$\pm$0.01 \\
NGC 5252     &  17.0$\pm$0.6 &  13.5$\pm$0.7 &  11.0$\pm$0.3 & 0.71$\pm$0.02 & 1.43$\pm$0.03 \\
NGC 5427     &  16.4$\pm$0.9 &  13.1$\pm$1.1 &  11.4$\pm$0.5 & 0.67$\pm$0.04 & 1.33$\pm$0.05 \\
NGC 5506     &   7.9$\pm$1.0 &   5.8$\pm$1.1 &   5.7$\pm$0.5 & 0.93$\pm$0.05 & 1.29$\pm$0.04 \\
NGC 5643     &   9.7$\pm$0.5 &   6.3$\pm$0.6 &   6.3$\pm$0.3 & 0.54$\pm$0.02 & 1.09$\pm$0.02 \\
NGC 5674     &  12.4$\pm$1.1 &  11.3$\pm$1.3 &   7.9$\pm$0.6 & 0.65$\pm$0.04 & 1.28$\pm$0.05 \\
NGC 5728     &  11.7$\pm$0.5 &   9.3$\pm$0.6 &   7.8$\pm$0.3 & 0.60$\pm$0.02 & 1.16$\pm$0.02 \\
NGC 5953     &  11.7$\pm$0.5 &   8.3$\pm$0.6 &   7.1$\pm$0.3 & 0.58$\pm$0.02 & 1.24$\pm$0.02 \\
NGC 6221     &   6.9$\pm$0.5 &   2.1$\pm$0.6 &   2.7$\pm$0.3 & 0.78$\pm$0.02 & 1.05$\pm$0.02 \\
NGC 6300     &  16.6$\pm$1.2 &   9.9$\pm$1.4 &   9.5$\pm$0.7 & 0.58$\pm$0.04 & 1.58$\pm$0.07 \\
NGC 6890     &  13.7$\pm$0.8 &   7.9$\pm$1.0 &   8.9$\pm$0.5 & 0.68$\pm$0.03 & 1.42$\pm$0.04 \\
NGC 7172     &  16.7$\pm$1.3 &  13.0$\pm$1.4 &  10.8$\pm$0.6 & 0.53$\pm$0.05 & 1.56$\pm$0.07 \\
NGC 7212     &  12.0$\pm$0.6 &   8.2$\pm$0.8 &   8.0$\pm$0.4 & 0.84$\pm$0.03 & 1.44$\pm$0.03 \\
NGC 7314     &  17.8$\pm$1.9 &  11.7$\pm$2.5 &  10.0$\pm$1.1 & 0.67$\pm$0.09 & 1.37$\pm$0.11 \\
NGC 7496     &   4.1$\pm$0.4 &   3.4$\pm$0.5 &   3.8$\pm$0.2 & 0.78$\pm$0.01 & 0.94$\pm$0.01 \\
NGC 7582     &   4.5$\pm$0.4 &   1.6$\pm$0.5 &   3.0$\pm$0.2 & 0.59$\pm$0.01 & 1.18$\pm$0.01 \\
NGC 7590     &  16.3$\pm$0.7 &  10.2$\pm$0.9 &  10.2$\pm$0.4 & 0.59$\pm$0.03 & 1.51$\pm$0.04 \\
NGC 7679     &   2.1$\pm$0.6 &   2.2$\pm$0.6 &   2.7$\pm$0.3 & 0.55$\pm$0.01 & 0.89$\pm$0.01 \\
NGC 7682     &  16.3$\pm$0.9 &  13.1$\pm$1.1 &  10.2$\pm$0.5 & 0.73$\pm$0.04 & 1.43$\pm$0.05 \\
NGC 7743     &  13.6$\pm$0.7 &   8.5$\pm$0.8 &   8.2$\pm$0.4 & 0.54$\pm$0.02 & 1.27$\pm$0.03 \\ \hline
MRK 0883     &   6.2$\pm$0.5 &   5.1$\pm$0.7 &   2.7$\pm$0.3 & 0.82$\pm$0.02 & 1.10$\pm$0.02 \\
NGC 1097     &   5.1$\pm$0.5 &   5.3$\pm$0.7 &   5.2$\pm$0.3 & 0.74$\pm$0.02 & 0.99$\pm$0.02 \\
NGC 4303     &  11.8$\pm$0.6 &   8.7$\pm$0.7 &   7.4$\pm$0.3 & 0.64$\pm$0.02 & 1.22$\pm$0.03 \\
NGC 4602     &   9.6$\pm$0.7 &   7.6$\pm$1.0 &   8.1$\pm$0.5 & 0.59$\pm$0.03 & 1.11$\pm$0.03 \\
NGC 7410     &   3.3$\pm$0.4 &   3.7$\pm$0.5 &   3.8$\pm$0.3 & 0.75$\pm$0.02 & 0.92$\pm$0.01 \\
ESO 108-G17  &   2.4$\pm$0.6 &   1.4$\pm$0.9 &   2.2$\pm$0.5 & 1.06$\pm$0.03 & 0.81$\pm$0.02 \\
NGC 1487     &   2.0$\pm$0.7 &   4.5$\pm$1.1 &   3.4$\pm$0.6 & 0.86$\pm$0.03 & 0.69$\pm$0.02 \\
NGC 2935     &  12.8$\pm$0.6 &  11.5$\pm$0.7 &   9.8$\pm$0.3 & 0.61$\pm$0.02 & 1.33$\pm$0.03 \\
NGC 3256     &   3.6$\pm$0.5 &   1.7$\pm$0.6 &   2.8$\pm$0.3 & 0.77$\pm$0.02 & 0.89$\pm$0.01 \\
NGC 2811     &  18.6$\pm$0.6 &  18.2$\pm$0.7 &  12.7$\pm$0.3 & 0.57$\pm$0.02 & 1.57$\pm$0.03 \\
NGC 3223     &  17.2$\pm$0.7 &  15.6$\pm$0.8 &  12.2$\pm$0.4 & 0.63$\pm$0.03 & 1.59$\pm$0.04 \\
NGC 3358     &  18.4$\pm$0.7 &  15.9$\pm$0.8 &  11.8$\pm$0.4 & 0.61$\pm$0.03 & 1.56$\pm$0.04 \\
NGC 3379     &  19.0$\pm$0.6 &  18.6$\pm$0.6 &  12.5$\pm$0.3 & 0.59$\pm$0.02 & 1.49$\pm$0.03 \\
NGC 4365     &  18.6$\pm$0.5 &  18.5$\pm$0.6 &  12.2$\pm$0.3 & 0.54$\pm$0.02 & 1.56$\pm$0.03 \\
\hline
\end{tabular}
\end{centering}
\caption{Spectral indices, computed with the definitions of Cid
Fernandes \etal (2004). Colours were corrected for Galactic
extinction.}
\label{tab:Stellar_Indices}
\end{table}

\subsection{Indirect measurements}

\label{sec:Synthetic_Indices}

Unfortunately, other useful tracers of the star-formation history such
as the $D_n(4000)$ and H$\delta_A$ indices (Balogh \etal 1999; Worthey
\& Otaviani 1997), cannot be measured directly for many of our
galaxies because of severe contamination by emission lines. An indirect
measurement of these indices can be performed over the model starlight
spectra described in \S\ref{sec:SpecSynthesis}.  Naturally, the
information contained in these ``semi-empirical'' indices is identical
to the one contained in the model spectra from which they are
measured. They are nevertheless useful for stellar population
diagnostics based on observable indices, such as the $D_n(4000)$
versus H$\delta_A$ diagram amply explored by Kauffmann \etal (2002,
2003) in their study of SDSS galaxies.

Figure \ref{fig:C3660_x_WK}b shows $W_K$ and $C_{3660}$ measured from
the deredenned synthetic spectra. The main difference with respect to
the observed version of this diagram (figure \ref{fig:C3660_x_WK}a) is
in the bottom-right points, which move upwards (to within the region
spanned by the BC03 models) due to the correction for intrinsic
reddening.  Objects in which a BLR is detected either in direct or
polarized light (ie, BLS2s, represented by filled circles) occupy a
characteristic region in this plot, close to the power-law + old stars
mixing line.  This is consistent with the existence of an FC implied
by the detection of a BLR.  This tendency is also evident, albeit less
clear, in the observed $W_K$-$C_{3660}$ diagram (figure
\ref{fig:C3660_x_WK}a).

\begin{figure}
\includegraphics[width=9cm]{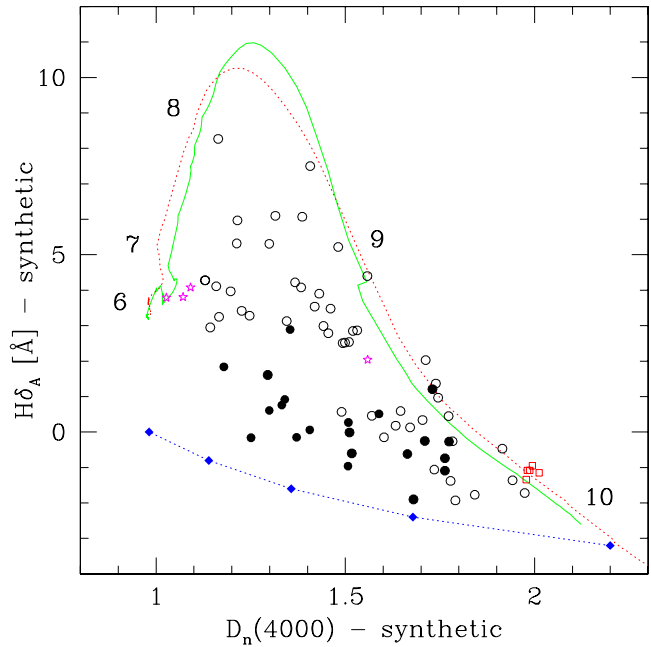}
\caption{$D_n(4000)$ versus H$\delta_A$ diagram. As in
Figure \ref{fig:C3660_x_WK}b, both indices were measured from the
synthetic spectra. The numbers around the BC03 SSP evolutionary
sequences mark the logarithm of the age (in yr). Symbols and lines as
in figure \ref{fig:C3660_x_WK}.}
\label{fig:D4000_x_Hdelta}
\end{figure}

Figure \ref{fig:D4000_x_Hdelta} shows the synthetic
$D_n(4000)$-H$\delta_A$ diagram for our sample.  As in figure
\ref{fig:C3660_x_WK}, the scatter of points in this plot reflects the
wide variety of stellar populations of Seyfert 2s, ranging from
systems dominated by young stars, to post-starbursts and older
systems, as well as mixtures of these populations. Again, objects
where an FC is known to be present from the detection of a BLR are
systematically off-set towards the power-law + old stars mixing line.
Note, however, that even these objects cannot be entirely explained in
terms of this simple two-components model.  An illustrative example in
this respect is Mrk 1210, with $D_n(4000) = 1.3$ and H$\delta_A = 1.6$
\AA. This galaxy has both a hidden BLR detected through
spectropolarimetry (Tran, Miller \& Kay 1992) and a very young
starburst, responsible for its WR bump (Storchi Bergmann, Cid
Fernandes \& Schmitt 1998; Joguet 2001). Other examples include Mrk
477 (Heckman \etal 1997; Tran \etal 1992), Mrk 463E
(Gonz\'alez-Delgado \etal 2001; Miller \& Goodrich
1990) and NGC 424 (Moran \etal 2000; Joguet 2001).

\section{Discussion}

\label{sec:Discussion}

\subsection{Statistics of star-formation in Seyfert 2s}

\begin{figure*}
\resizebox{\textwidth}{!}{\includegraphics{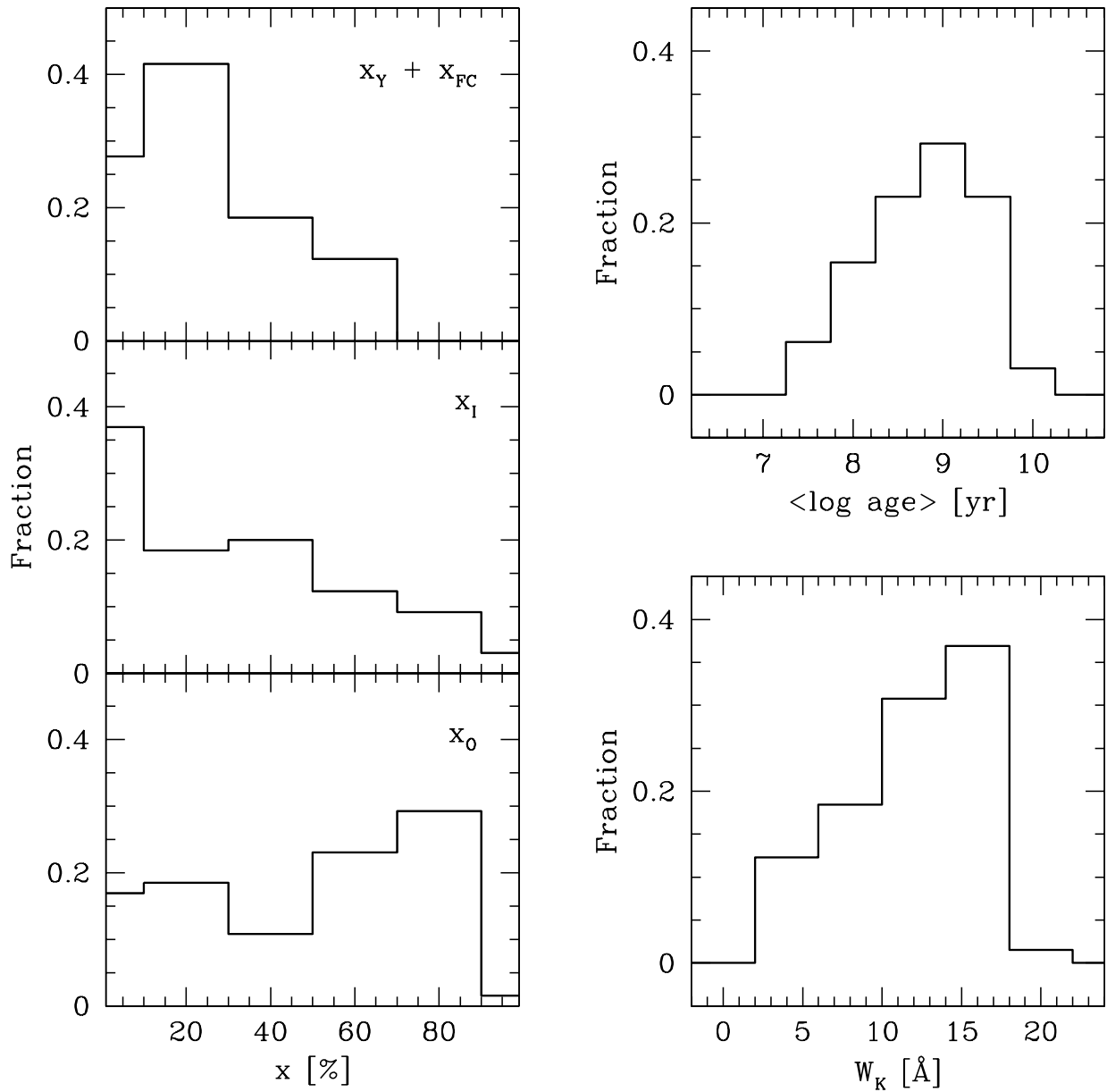}}
\caption{Distributions of the $x_{Y/FC}$, $x_I$ and $x_O$ fractions,
$W_K$ and the flux weighted mean stellar age inferred from the
synthesis.  Only Seyfert 2s were included in these histograms.  }
\label{fig:histograms}
\end{figure*}

Figure \ref{fig:histograms} shows histograms of the $x_{Y/FC}$, $x_I$
and $x_O$ components for the 65 Seyfert 2s in the sample, as well as
the distributions of $W_K$ and $\overline{\log t}$, the flux weighted
mean stellar age (computed according to its definition in Cid
Fernandes, Rodrigues Lacerda \& Le\~ao 2003). These histograms
illustrate the frequency of recent star-formation in Seyfert 2s.

There are several possible definitions of ``recent star-formation'',
either in terms of observed quantities or model parameters.  For
instance, 31\% of the objects have $W_K < 10$ \AA, a rough dividing
line between composite starburst + Seyfert 2 systems and those
dominated by older populations according to CF01. This frequency
increases to 51\% using a $W_K < 12$ \AA\ cut.  In terms of
$x_{Y/FC}$, we find that nearly half (46\%) of the sample has
$x_{Y/FC} > 20\%$, while for about one-third this fraction exceeds
30\%. Old stars, although present in all galaxies, only dominate the
light ($x_O > 70\%$) in 31\% of the cases.  In terms $\overline{\log
t}$, we find that 25 out of our 65 galaxies (38\%) have
$\overline{\log t} < 8.5$, ie, mean ages smaller than 300 Myr.  In
summary, depending on the criterion adopted, between 1/3 and 1/2 of
Seyfert 2s have experienced significant star-formation in the past few
hundred Myr.  These numbers are similar to those derived by J01 by
means of a visual characterization of stellar populations for the same
sample and with those obtained by Storchi-Bergmann \etal (2001) for a
sample $\sim$ half as big as ours.

Naturally, one would like to compare the incidence of recent
star-formation in Seyfert 2s to that in other types of nuclei. A
particularly interesting comparison to be made is that between Seyfert
2s and LINERs, given that these two classes were once thought to
belong to the same family (Ferland \& Netzer 1983; Halpern \& Steiner
1993). We have too few LINERs in our sample to perform this
comparison, but we may borrow from the results of Cid Fernandes \etal
(2004) and Gonz\'alez Delgado \etal (2004), who have recently surveyed
the stellar populations of LINERs and LINER/HII Transition Objects
using spectroscopic data practically identical to the one used in this
paper and a similar method of analysis. Their results are {\it very
different} from ours. For instance, while we find that over 2/3 of
Seyfert 2s have $x_{Y/FC} > 10\%$, most LINERs have $x_{Y/FC} < 10\%$,
and are often dominated by old stars.  Still according to these
authors, a large fraction of Transition Objects harbour significant
numbers of $10^8$--$10^9$ yr stars, similar to those found in
$x_I$-dominated Seyfert 2s (eg, ESO 362-G08 and Mrk 1370, figures
\ref{fig:Fit_j008} and \ref{fig:Fit_j026}).  Their $x_{Y/FC}$
strength, however, is only 6\% on average, compared to 23\% for our
Seyfert 2s.

Thus, a substantial fraction of nearby Seyfert 2s live in much younger
stellar environments than LINERs or Transition Objects, which suggests
an evolutionary sequence. Given that the later objects are known to
harbour less luminous AGN and have different gas excitation conditions
than Seyferts, if confirmed, this would imply a substantial evolution
of the AGN itself---in parallel with the evolution of stars in its
surroundings.

\subsection{Aperture effects}

\begin{figure}
\includegraphics[width=9cm]{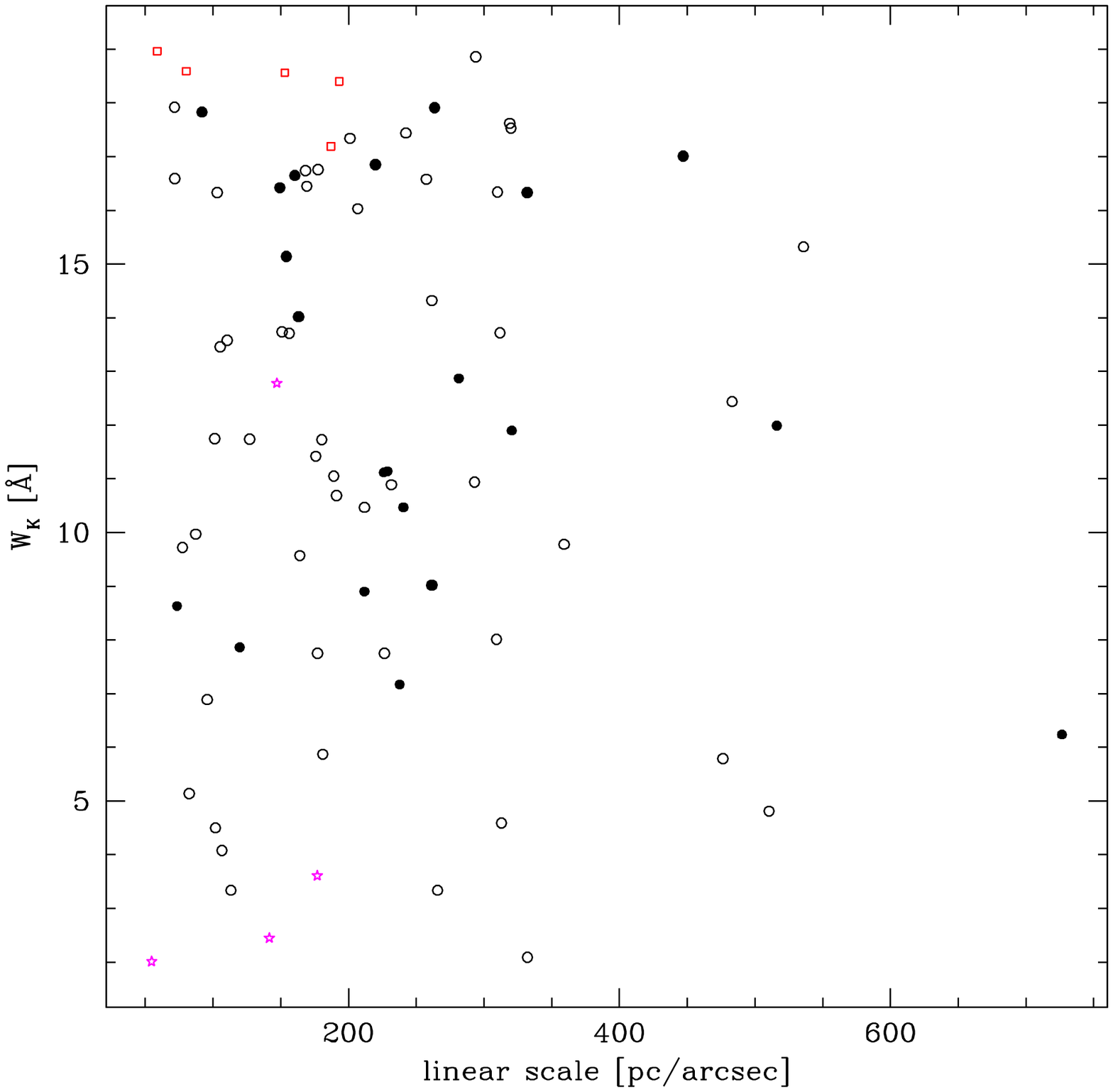}
\caption{Projected linear scale against the equivalent width of CaII
K. Symbols as in figure \ref{fig:YIO_triangle}.}
\label{fig:distance_effect}
\end{figure}

Spectra collected through apertures of fixed angular size sample
distance-dependent linear scales, which may introduce systematic
effects on the derived stellar population mixtures.  In order to
investigate whether such potential biases affect our results in some
systematic way we plot in figure \ref{fig:distance_effect} the
equivalent width of CaII K against the pc/arcsec scale. (Recall that
our observations cover an area of $\sim 1$ arcsec$^2$.) No systematic
trend is observed between this stellar population tracer and the
linear dimension of the regions sampled by our data. The same
conclusion holds for the $\vec{x}$ components, none of which
correlates with distance either.  The absence of systematic trends
indicates that our analysis is not affected by aperture effects.

\subsection{Relation to host galaxy morphology}

\label{sec:Morphology}

A galaxy's morphology carries important information for the study of
star formation history in galaxies (eg., Kennicutt 1998). From early
to late types, galaxy colours become bluer and star formation rates
higher.  Storchi-Bergmann \etal (2001) studied the relationship
between nuclear stellar population and morphology in a sample of 35
Seyfert 2s, 13 in common with our sample. They found that the fraction
of galaxies with recent nuclear/circumnuclear starburst increases
along the Hubble sequence and that there is a relation between the
presence of a nuclear starburst and a late ``inner Hubble type'' from
the HST classifications of Malkan, Gorjian \& Tam (1998).

\begin{figure*}
\resizebox{\textwidth}{!}{\includegraphics{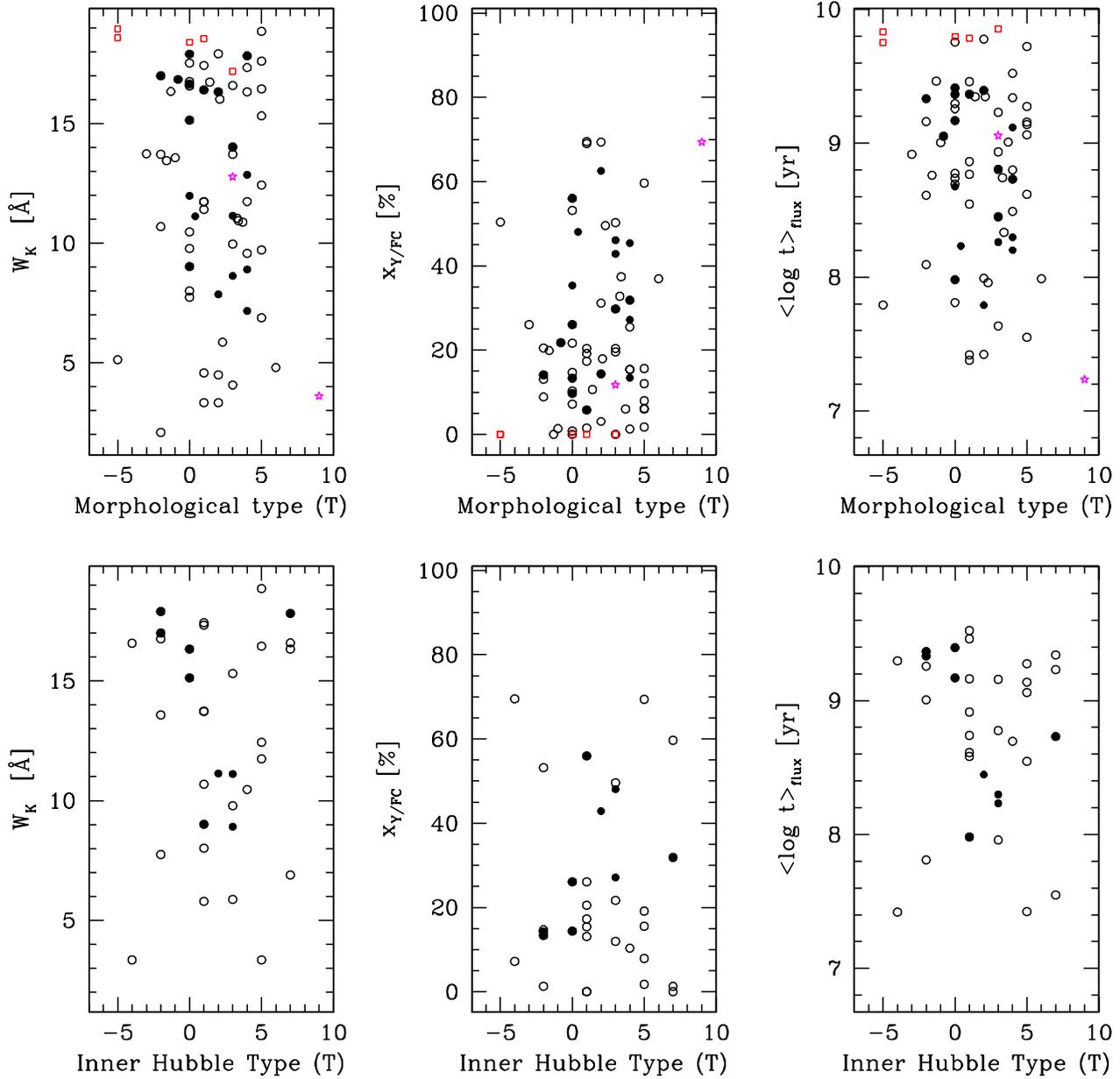}}
\caption{Distribution of morphological Hubble type for the host
galaxies (top panels) and the inner region (bottom panels) against
stellar population properties: W$_K$ (left), $x_{Y/FC}$ (middle) and
the mean age (right). Symbols as in figure \ref{fig:YIO_triangle}.}
\label{fig:Morphology}
\end{figure*}

In order to check this result in our larger sample, we have collected
morphological information on the host galaxies from RC3, and the inner
Hubble types from Malkan \etal (1998).  In figure~\ref{fig:Morphology}
we plot three of our stellar population tracers (W$_K$, x$_{Y/FC}$,
and mean age) against the RC3 and inner morphological types (later
types have larger numbers).  There is no correlation in our sample
between (optical) morphology and nuclear star formation.  Since the
HST snapshot survey of Malkan \etal (1998) was taken using the F606W
filter which is sensitive to dust extinction and H$\alpha$ emission,
we searched the HST/NICMOS archive data and found images for 36 of our
galaxies taken with the F160W filter. No clear trend between
morphological type and nuclear stellar population is found in the
near-IR images either.

Finally, following Storchi-Bergmann \etal we also looked for
companions to our Seyfert 2s according to angular separation, radial
velocity, and apparent magnitude. We find that 40\% of the galaxies
with $W_K < 10$ \AA\ and 47\% of those with $W_K>10$ \AA\ show the
presence of companions, compared respectively with 42\% and 47\% that
do not have near neighbours.

Hence, contrary to the findings of Storchi-Bergmann \etal (2001) we
find no correlation between the nuclear stellar populations of Seyfert
2 galaxies and either morphological types or environments.  Since
there is a substantial overlap between our samples, the discrepancy is
most likely statistical. Given the importance of the result, however,
it would be worthwhile to repeat the study in an even larger sample of
galaxies.

\section{Summary}

\label{sec:Conclusions}

We have presented a study of the stellar population in the central
$\sim 50$--500 pc of a large, well defined and homogeneous sample of
Seyfert 2s from the atlas of J01. Our main results may be summarized
as follows:

\begin{enumerate}

\item We have developed a spectral synthesis code which decomposes an
observed spectrum as a sum of simple stellar populations, represented
by state-of-the-art evolutionary synthesis models, plus an AGN
continuum.  Unlike previous studies, which use only a few absorption
equivalent widths and colours (eg Bica 1988; Schmitt \etal 1999; Cid
Fernandes \etal 2001a,b), we now synthesize the whole observed
spectrum. The method produces an estimate of the star-formation
history in terms of flux ($\vec{x}$) or mass ($\vec{\mu}$) fractions
associated to each component in the spectral base, as well as
extinction and velocity dispersions. As in other population synthesis
studies, the accuracy of our method is limited by noise, spectral
coverage, and intrinsic degeneracies of stellar populations. These
limitations are not critical as long as one does not attempt an
over-detailed description of the SFH.

\item The synthesis method was applied to 3500--5200 \AA\ spectra of
65 Seyfert 2s and 14 other galaxies from the J01 atlas. The synthesis
produces excellent fits of the observed spectra, with typical flux
residuals of 2--5\%. The star formation history deduced from these
fits is remarkably varied among Seyfert 2s. Young starbursts,
intermediate age, and old populations all appear in significant and
mixed amounts.

\item The stellar velocity dispersions obtained from the synthesis are
in good agreement with measurements in the literature for galaxies in
common. This is important since this information gives us a handle on
the black-hole mass (Ferrarese \& Merrit 2000; Gebhart \etal 2000).

\item The starlight-subtracted spectra were used to investigate the
presence of weak broad emission features which are hard to detect in
the total spectrum. This analysis revealed the signatures of WR stars
in 3 (maybe 5) Seyfert 2s, all of which have large young stellar
populations as deduced by the synthesis.

\item The analysis of the ``pure-emission'' spectra further allowed
the detection of a weak BLR-like component under H$\beta$ in several
Seyfert 2s. For most of these objects independent spectropolarimetry
data reveals a type 1 spectrum. We are thus detecting in direct,
non-polarized light the scattered BLR predicted by the unified model.

\item The detection of a BLR either in direct or polarized light
implies that an associated FC should also be present in our spectrum.
Indeed, the FC strengths obtained by the spectral decomposition for
the 17 Seyfert 2s in our sample which present such evidence are
substantially larger than for nuclei with no indications of a BLR.
The fraction of these ``broad line Seyfert 2s'' increases
systematically with increasing $x_{FC}$.

\item Dusty young starbursts can also appear disguised as an
AGN-looking continuum. We have identified several cases where this
indeed happens. These Seyfert 2s tend to have large $x_{FC}$ {\it and}
$x_Y$ fractions. The information on the existence of a BLR, and thus
of a true FC component, helps breaking, at least partly, this
starburst-FC degeneracy. Whenever the spectral synthesis identifies a
strong ($\ga 20\%$) FC but no BLR is detected either in the residual
or polarized spectra, the FC is most likely dominated by a dusty
starburst rather than scattering of a hidden AGN.

\item Stellar indices were measured both from the observed and
synthetic spectra to provide a more empirical characterization of the
stellar content of our galaxies and facilitate comparisons with
independent studies. These indices, which include $W_K$, the 3660/4020
colour, $D_n(4000)$ and H$\delta_A$, confirm the heterogeneity of
nuclear stellar populations in Seyfert 2s.

\item Between 1/3 and 1/2 of nearby Seyfert 2s have experienced
significant star formation in the recent past. Thus, as a class,
Seyfert 2 nuclei live in a much younger stellar environment than
LINERs and normal galactic nuclei.

\item We find no significant correlation between the host morphology
(as deduced both from ground based and HST optical and near-IR images)
and stellar population in Seyfert 2s. The presence of companions does
not seem to correlate with the stellar populations either.

\end{enumerate}

We have shown that using modern population synthesis models it is
possible to obtain meaningful information about the history of star
formation in objects for which the spectral data are dominated by
strong emission lines. The road to detailed quantitative
investigations of the relationship between star-formation and AGN is
now open. We will be traveling along it in our forthcoming papers.

\section*{ACKNOWLEDGMENTS}

QG thanks the hospitality of UFSC and ESO-Santiago and the support
from CNPq and the National Natural Science Foundation of China under
grants 10103001 and 10221001 and the National Key Basic Research
Science Foundation (NKBRSG19990754). Partial support from CNPq and
PRONEX are also acknowledged. We also thank the anonymous referee for
his/her careful reading and constructive criticism of the original
manuscript.


\begin{thebibliography}{99}

\bibitem[]{} Antonucci R., 1993, ARA\&A, 31, 473
\bibitem[]{} Antonucci R., Miller J.S., 1985, ApJ, 297, 621
\bibitem[]{} Aretxaga I., Terlevich E., Terlevich R., Cotter G., Diaz
A., 2001, MNRAS, 325, 636
\bibitem[]{} Aretxaga I., Joguet B., Kunth D., Melnick J., Terlevich
R., 1999, ApJ, 519, L123
\bibitem[]{} Balogh, M.~L., Morris, S.~L., Yee, H.~K.~C., Carlberg,
R.~G., \& Ellingson, E.\ 1999, ApJ, 527, 54
\bibitem[]{} Bica E., Alloin D.,1986, A\&A, 162, 21 
\bibitem[]{} Boisson C., Joly M., Moultaka J., Pelat D., Serote Roos
M., 2000, A\&A, 357, 850
\bibitem[]{} Bruzual G., Charlot S., 2003, MNRAS, 344, 1000
\bibitem[]{} Cardelli J. A., Clayton G. C., Mathis J.S., 1989, ApJ, 345, 245
\bibitem[]{} Chabrier G., 2003, PASP, 115, 763
\bibitem[Chan, Mitchell, \& Cram(2003)]{2003MNRAS.338..790C} Chan,
B.~H.~P., Mitchell, D.~A., \& Cram, L.~E.\ 2003, MNRAS, 338, 790
\bibitem[]{} Cid Fernandes R., Terlevich R., 1995, MNRAS, 272, 423
\bibitem[]{} Cid Fernandes R., Storchi-Bergmann T., Schmitt H. R.,
1998, MNRAS, 297, 579
\bibitem[]{} Cid Fernandes, R., Heckman, T., Schmitt, H., Gonzalez
Delgado, R.~M., \& Storchi-Bergmann, T.\ 2001a, ApJ, 558, 81
\bibitem[]{} Cid Fernandes R., Sodr\'e L., Schmitt H. R., Le\~ao
J. R. S., 2001b, MNRAS, 325, 60
\bibitem[]{} Cid Fernandes R., Leao J. Lacerda R. R., 2003, MNRAS,
340, 29
\bibitem[]{} Cid Fernandes R., Gonz\'alez Delgado R. M., Schmitt H.,
Storchi-Bergmann T., Pires Martins L., P\'erez E., Heckman T.,
Leitherer C., Schaerer D. 2004, ApJ, 605, 105
\bibitem[]{} Colina L., Gonzalez Delgado R., Mas-Hesse J. M.,
Leitherer C., 2002, ApJ, 579, 545
\bibitem[]{} Colina L., Cervino, M., Gonzalez Delgado, R.M. , 2003,
ApJ, 593, 127
\bibitem[]{} Della Ceca R., Pellegrini S., Bassani L., et al., 2001,
A\&A, 375, 781
\bibitem[]{} Dessauges-Zavadsky M., Pindao M., Maeder A., Kunth D.,
2000 A\&A, 355, 89
\bibitem[]{} de Vaucouleurs G., de Vaucouleurs A., Corwin H.G., Buta
R.J., Paturel G., Fouque P., 1991, Springer-Verlag: New York, Third
Reference Catalogue of Bright Galaxies (RC3)
\bibitem[]{} Ferland G.J., Netzer H., 1983, ApJ, 264, 105
\bibitem[]{} Ferrarese L., Merritt D., 2000, ApJ, 539, L9
\bibitem[]{} Gebhardt K., et al., 2000, ApJ, 539, L13
\bibitem[]{} Goncalves A., Veron-Cetty M., Veron P., 1999, A\&AS, 135,
437
\bibitem[]{} Gonzalez Delgado R., Heckman T., Leitherer C., et al.,
1998, ApJ, 505, 174
\bibitem[]{} Gonzalez Delgado R., Leitherer C., Heckman T., 1999,
ApJS, 125, 489
\bibitem[]{} Gonzalez Delgado R., Heckman T., Leitherer C., 2001, ApJ,
546, 845
\bibitem[]{} Gonzalez Delgado R., Cid Fernandes R., P\'erez E., Pires
Martins L., Storchi-Bergmann T., Schmitt H., Heckman T., Leitherer C.,
2004 ApJ, 605, 127
\bibitem[]{} Gu, Q.~S., Huang, J.~H., de Diego, J.~A., Dultzin-Hacyan,
D., Lei, S.~J., \& Ben{\'{\i}}tez, E.\ 2001, A\&A, 374, 932
\bibitem[]{} Gu Q., Huang J., 2002, ApJ, 579, 205
\bibitem[]{} Heckman T., Krolik J., Meurer G., Calzetti, D., Kinney,
A., Koratkar, A., Leitherer, C., Robert, C., Wilson, A., 1995, ApJ,
452, 549
\bibitem[]{} Heckman, T.~M., Gonzalez-Delgado, R., Leitherer, C.,
Meurer, G.~R., Krolik, J., Wilson, A.~S., Koratkar, A., \& Kinney, A.\
1997, ApJ, 482, 114
\bibitem[]{} Halpern J.P., Steiner J.E., 1983, ApJ, 269, L37
\bibitem[]{} Jimenez-Bailon E., Santos-Lleo M., Mas-Hesse J. M.,
Guainazzi M., 2003, ApJ, 593, 127
\bibitem[]{} Joguet B., 2001, PhD thesis, Institut D'Astrophysique de
Paris
\bibitem[]{} Joguet B., Kunth D., Melnick J., Terlevich R., Terlevich
E., 2001, A\&A, 380, 19 (J01)
\bibitem[]{} Kauffmann G., Heckman T., White S., et al., 2003, MNRAS,
341, 33
\bibitem[]{} Kauffmann G., Heckman T., Tremonti C., et al., 2003,
MNRAS, in press (astro-ph/0304239)
\bibitem[]{} Kennicutt R. C. Jr., 1998, ARA\&A, 36, 189
\bibitem[]{} Khachikian E.Y., Weedman D.W., 1974, ApJ, 192, 581
\bibitem[]{} Koski A.T., 1978, ApJ, 223, 56
\bibitem[]{} Le Borgne J.-F. et al., 2003, A\&A, 402, 433
\bibitem[]{} Levenson N., Cid Fernandes R., Weaver K., Heckman T.,
Storchi-Bergmann T., 2001, ApJ, 557, 54
\bibitem[]{} Lira P., Ward M., Zezas A., Alonso-Herrero A., Ueno S.,
2002, MNRAS, 330, 259
\bibitem[]{} Lumsden, S. L., Alexander, D. M., Hough, J. H., 2004,
MNRAS, 348, 1451 
\bibitem[]{} Maiolino R., et al., 2003, MNRAS, 344, L59
\bibitem[]{} Malkan M.A., Gorjian V., Tam R., 1999, ApJS, 117, 25
\bibitem[]{} Mayya, Y.~D., Bressan, A., Rodr{\'{\i}}guez, M., Valdes,
J.~R., \& Chavez, M.\ 2004, ApJ, 600, 188
\bibitem[]{} Melnick J., Gopal-Krishna, Terlevich R., 1997, A\&A, 318,
337
\bibitem[]{} Miller J.S., Goodrich R.W., 1990, ApJ, 355, 456
\bibitem[]{} Moran E., Barth A., Kay L., Filippenko A. V., 2000, ApJ,
540, L73
\bibitem[]{} Moultaka J., Pelat D., 2000, MNRAS, 314, 409
\bibitem[]{} Nelson C., Whittle M., 1995, ApJS, 99, 67
\bibitem[]{} Osterbrock D. E. 1984, Quart. J. R. Astron. Soc., 25, 1
\bibitem[]{} Pelat D., 1997, MNRAS, 284, 365
\bibitem[]{} Schaerer D., Contini T., Kunth D., 1999, A\&A, 341, 399
\bibitem[]{} Schlegel D., Finkbeiner D., Davis M., 1998, ApJ, 500, 525
\bibitem[]{} Schmidt A. A., Copetti M. V. F., Alloin D., Jablonka
P. 1991, MNRAS, 303,173
\bibitem[]{} Schmitt H., Storchi-Bergmann T., Cid Fernandes R., 1999,
MNRAS, 303,74
\bibitem[]{} Selman F., Melnick J., Bosch G., Terlevich R., 1999,
A\&A, 341, 98
\bibitem[]{} Serote Roos, M., Boisson, C., Joly, M., Ward, M. J.,
1998, MNRAS, 301, 1
\bibitem[]{} Storchi-Bergmann T., Cid Fernandes R., Schmitt H., 1998,
ApJ, 501, 94
\bibitem[]{} Storchi-Bergmann T., Raimann D., Bica E., Fraquelli H.A.,
2000, ApJ, 544, 747
\bibitem[]{} Storchi-Bergmann T. et al., 2003, ApJ, 598, 956
\bibitem[]{} Tremonti C., 2003, PhD thesis, Johns Hopkins University
\bibitem[]{} Tran H.D.,  Miller J., Kay L, 1992, ApJ, 397, 452
\bibitem[]{} Tran H.D., 1995a, ApJ, 440, 578
\bibitem[]{} Tran H.D., 1995b, ApJ, 440, 597
\bibitem[]{} Tran H.D., 2001, ApJ, 554, L19
\bibitem[]{} Terlevich E., Diaz A. I., Terlevich R., 1990, MNRAS, 242,
271.
\bibitem[]{} Wills, K.~A., Tadhunter, C.~N., Robinson, T.~G., \&
Morganti, R.\ 2002, MNRAS, 333, 211
\bibitem[]{} Worthey G., 1994, ApJS, 94, 687
\bibitem[]{} Worthey G., Ottaviani D. L., 1997, ApJS, 111, 377


\end{thebibliography}
\end{document}